\documentclass[12pt, draftclsnofoot, onecolumn]{IEEEtran}

\usepackage[utf8]{inputenc}
\usepackage{cite}
\usepackage{graphicx}
\usepackage{amsmath}
\usepackage{mathrsfs,amsthm,amsmath,amssymb}
\usepackage{textcomp}
\usepackage{graphicx,multirow}
\usepackage{balance}
\usepackage{cite}
\usepackage{float}
\usepackage{gensymb}
\usepackage{eso-pic}
\usepackage{algorithm}
\usepackage{algorithmic}
\usepackage[hidelinks]{hyperref}
\usepackage{academicons}
\usepackage{xcolor}
\usepackage{pdfpages}

\makeatletter
\newcommand\subparagraph{%
  \@startsection{subparagraph}{5}
  {\parindent}
  {3.25ex \@plus 1ex \@minus .2ex}
  {-1em}
  {\normalfont\normalsize\bfseries}}
\makeatother
\usepackage{titlesec}
\let\subparagraph\relax

\usepackage{titlesec}
\usepackage{scalerel}
\usepackage{tikz}
\usepackage{subcaption}

\usepackage[english]{babel}

\newtheorem{theorem}{Theorem}

\newcommand{\mb}{\mathbf}

\DeclareMathOperator{\tr}{trace}
\DeclareMathOperator{\diag}{diag}
\DeclareMathOperator{\vect}{vec}

\usetikzlibrary{svg.path}

\definecolor{orcidlogocol}{HTML}{A6CE39}
\tikzset{
  orcidlogo/.pic={
    \fill[orcidlogocol] svg{M256,128c0,70.7-57.3,128-128,128C57.3,256,0,198.7,0,128C0,57.3,57.3,0,128,0C198.7,0,256,57.3,256,128z};
    \fill[white] svg{M86.3,186.2H70.9V79.1h15.4v48.4V186.2z}
                 svg{M108.9,79.1h41.6c39.6,0,57,28.3,57,53.6c0,27.5-21.5,53.6-56.8,53.6h-41.8V79.1z M124.3,172.4h24.5c34.9,0,42.9-26.5,42.9-39.7c0-21.5-13.7-39.7-43.7-39.7h-23.7V172.4z}
                 svg{M88.7,56.8c0,5.5-4.5,10.1-10.1,10.1c-5.6,0-10.1-4.6-10.1-10.1c0-5.6,4.5-10.1,10.1-10.1C84.2,46.7,88.7,51.3,88.7,56.8z};
  }
}

\newcommand\orcidicon[1]{\href{https://orcid.org/#1}{\mbox{\scalerel*{
\begin{tikzpicture}[yscale=-1,transform shape]
\pic{orcidlogo};
\end{tikzpicture}
}{|}}}}

\DeclareMathOperator{\st}{s.t.}
\DeclareMathOperator{\re}{Re}

\graphicspath{ {Images/} }
\IEEEoverridecommandlockouts

\begin{document}
\setlength{\parskip}{5pt}
\setlength{\abovedisplayskip}{5pt}
\setlength{\belowdisplayskip}{5pt}

\title{Low-Complexity Beamforming Design for IRS-Aided NOMA Communication System with Imperfect CSI}

\author{Yasaman~Omid$^\text{\orcidicon{0000-0002-5739-8617}}$, 
S. M. Mahdi~Shahabi$^\text{\orcidicon{0000-0003-2873-4156}}$, Cunhua~Pan$^\text{\orcidicon{0000-0001-5286-7958}}$,\\ 
Yansha~Deng$^\text{\orcidicon{0000-0003-1001-7036}}$, 
Arumugam~Nallanathan$^\text{\orcidicon{0000-0001-8337-5884}}$,~\IEEEmembership{Fellow,~IEEE}
 
 \thanks{Yasaman Omid, Cunhua Pan  and Arumugam Nallanathan  are with the School of Electronic Engineering and Computer Science, Queen Mary University of London, U.K. (e-mail: y.omid@qmul.ac.uk; c.pan@qmul.ac.uk; a.nallanathan@qmul.ac.uk).} 
 
%  \thanks{Seyyed MohammadMahdi Shahabi is with the Department of Electrical Engineering, K. N. Toosi University of Technology, Tehran, Iran (e-mail: shahabi@ee.kntu.ac.ir).}

\thanks{S. M. Mahdi Shahabi and Yansha Deng are with the Department of Engineering, Kings College London, U.K. (e-mail: mahdi.shahabi@kcl.ac.uk;  yansha.deng@kcl.ac.uk).}
}

\maketitle
\begin{abstract}
Intelligent reflecting surface (IRS) as a promising technology rendering high throughput in future communication systems is compatible with various communication techniques such as non-orthogonal multiple-access (NOMA). In this paper, the downlink transmission of IRS-assisted NOMA communication is considered while undergoing imperfect channel state information (CSI). Consequently, a robust IRS-aided NOMA design is proposed by solving the sum-rate maximization problem to jointly find  the optimal beamforming vectors for the access point  and the passive reflection matrix for the IRS, using the penalty dual decomposition (PDD) scheme. This problem can be solved through an iterative algorithm, with closed-form solutions in each step, and it is shown to have very close performance to its upper bound obtained from perfect CSI scenario. We also present a trellis-based method  for optimal discrete phase shift selection of IRS which is shown to outperform the conventional quantization  method.  Our results show that the proposed algorithms, for both continuous and discrete IRS, have very low computational complexity compared to other schemes in the literature. Furthermore, we conduct a performance comparison from achievable sum-rate standpoint between IRS-aided NOMA and IRS-aided orthogonal multiple access (OMA), which demonstrates   superiority of NOMA compared to OMA in case of a tolerated channel uncertainty.         

\end{abstract}
\begin {IEEEkeywords}
Intelligent Reflecting Surface, Robust Design,  NOMA, Penalty Dual Decomposition, Trellis.  
\end{IEEEkeywords}
\section{Introduction}
\label{intro}
\IEEEPARstart{O}{ne} of the most recognized multiple-access techniques for future communication systems is the non-orthogonal multiple access (NOMA). The basic idea {behind} NOMA is to serve multiple users on each orthogonal resource block. These  resource blocks could be orthogonal bandwidths, different time steps, or orthogonal spacial directions in case the system contains multiple-antenna access points (AP). In the orthogonal multiple access (OMA) techniques, each resource block is dedicated to only one user, while in NOMA, multiple users are served on the same resource block. Thus, the spectral efficiency of the system is {potentially} improved by using NOMA by a factor of $K$ compared to OMA systems, $K$ being the number of users served on each orthogonal resource block.  
% \textcolor{\omissioncolor}{OMA, on the other hand, is divided into the three categories of spatial-division multiple-access (SDMA), frequency-division multiple-access (FDMA) and time-division multiple-access (TDMA), depending on which orthogonal resource block we use. (This is a basic statement that should be removed)}    
Although it might seem that use of NOMA is always favorable, the application of NOMA is  limited to the case where the direction of users' channel vectors are similar \cite{Chen2016OnDownlink}. 
% \textcolor{\revisioncolor}{(You mentioned SDMA many times as an alternative to NOMA. However, you didn't analyze this method in the Numerical analysis. So, from this point of view, the paper suffers from lack of consistency. You should either analyze SDMA in the numerical section or append TDMA and FDMA in the introduction section rather than SDMA.)} 
Therefore, to broaden the application of NOMA and to improve its performance, intelligent reflecting surfaces (IRS) can be utilized to manipulate the direction of the users' channel vectors \cite{Ding2020ATransmittion}. 

IRS is one of the most promising performance-enhancement technologies for next-generation wireless networks to improve both the spectral efficiency and the energy efficiency  of the wireless communication systems \cite{Pan2020ReconfigurableDirections}. An IRS is basically a planar {array} which consists of a large number of passive phase shifters that reconfigure the wireless propagation environment by inducing phase shifts on the impinging electromagnetic waves \cite{Pan2019}, \cite{CunhuaNEW}.
It has been shown that IRS is compatible with many communication techniques, e. g. millimeter-wave, {Terahertz communications}, physical layer security, simultaneous wireless information and power transfer (SWIPT), unmanned aerial
vehicle networks (UAV), MIMO systems and finally, NOMA \cite{ZhangReflectionsReflectors2019,Robust3,WuWeighted2019,ZuoIntelligentSystems2020}. Most papers assume that the IRS lacks active elements, thus it is incapable of pilot-based channel estimation. 
Hence, in most IRS-assisted system designs, it is assumed that the cascaded AP-IRS-user channels are available at the AP, because the estimation of IRS-related channels separately is difficult. Furthermore, in some recent works, the effects of channel estimation error on the performance of the IRS is considered and robust designs have been proposed to cope with this issue \cite{Omiddd1,Robust1,Robust2}. Thus, in this paper, we attempt to design a robust IRS-aided NOMA communication system.   

\subsection{Prior Work Regarding IRS-Aided NOMA}
The application of  IRS for NOMA communication has only recently been noticed by some researchers. In \cite{Ding2020ATransmittion}, the authors proposed a simple IRS-assisted NOMA downlink transmission. At first, they employed { spatial-division  multiple-access  (SDMA)} at the AP to generate orthogonal beams, by using the spacial directions of the nearby users' channels. Then, the IRS-assisted NOMA was used to ensure that additional cell-edge users can also be served on these beams by aligning the cell-edge users effective channels with the predetermined beams. 
In \cite{Zheng2020IntelligentOMA?}, the performance comparison between NOMA and two types of OMA, namely {frequency-division multiple-access (FDMA)} and {time-division multiple-access (TDMA)} was studied in an IRS-aided downlink communication network. The transmit power minimization problem was solved for discrete phase shifters at the IRS. It was shown that TDMA is always more power-efficient than FDMA, but, the power consumption of TDMA compared to NOMA depends upon the target rate and the location of the users. The authors in \cite{Hou2020ReconfigurableNetworks} investigated the spectral efficiency improvement of NOMA networks by presenting an IRS-aided single-input single-output (SISO) network. To do so, they attempted to enhance the performance of the user with the best channel gain, while all other users depended on the IRS. 
The authors in \cite{Fu2019IntelligentNetworks} considered the downlink transmit power minimization problem for the IRS-aided NOMA system. They addressed the resulting intractable non-convex bi-quadratic problem by solving the non-convex quadratic problems alternatively, and to solve the non-convex quadratic problems, they employed a difference-of-convex (DC) programming algorithm.  
% In \cite{ZhangRobustNetworks2020}, the security of the IRS-assisted NOMA communication system was addressed in presence of an eavesdropper with imperfect channel estimation, while the legitimate users' CSI was perfectly known. This paper presents a robust beamforming method by using artificial noise to guarantee the security of communication in IRS-assisted NOMA \textcolor{\inquirycolor}{ (Is your work related to security?! If so, keep this part, otherwise, remove it)}. 
In \cite{MuExploitingOptimization2020}, for the first time, the sum-rate maximization problem was addressed  for a MISO IRS-aided NOMA system in the downlink transmission. By using the alternating optimization technique, they designed the passive phase shifts of the IRS and the active AP beamforming vector, alternatively, for two cases of ideal and non-ideal IRS phase shifters. In \cite{ZengSumNOMA2021}, the authors aimed to maximize the sum-rate in the uplink transmission of an IRS-aided NOMA system, and they proposed a semi-definite relaxation (SDR)-based algorithm that reached near-optimal solutions. They showed that the IRS-aided NOMA outperforms the IRS-aided OMA in terms of sum-rate.  
The Authors in \cite{ChenDownlinkOma2021}, study both uplink and downlink transmission of IRS-aided NOMA and OMA systems. Other papers in  the IRS-aided NOMA communication are listed here \cite{Wang2020OnSystems,Sena2020WhatAccess,Fang2020EnergyNetworks,Ding2020OnNOMA,Li2019JointNetworks}.

\subsection{Motivations and Contributions}
All of the aforementioned researches have considered availability of perfect CSI, while due to the passive nature of the IRS, channel acquisition in IRS-aided systems is quite challenging, especially for the IRS-related channels. Although, some papers have presented robust beamforming designs for IRS-aided MIMO systems  \cite{Omiddd1,Cascadedest1,Robust1,Zhou2020,Robust2}, to the best of our knowledge, no one has ever considered the effect of channel uncertainty in IRS-assisted NOMA and IRS-aided OMA systems.
To address this issue, in this paper a robust design for IRS-assisted NOMA communication system is {devised}. Motivated by \cite{Omiddd1}, the sum-rate maximization problem is reformulated into a tractable optimization problem using the penalty dual decomposition (PDD) scheme \cite{Pddmain1} and by applying the block successive upper bound minimization/maximization (BSUM) method \cite{Razaviyaynopt}, the problem is solved through an iterative algorithm. The solution in each iteration would be  closed-form which results in very low computational complexity. \textcolor{black}{ We consider two cases of continuous and discrete phase shifters in the IRS. In the literature, the authors either used exhaustive search to find optimal discrete phase shifts for the IRS, which requires huge computational complexity, or  simply quantized the obtained solution for a continuous IRS, which increases performance loss. In this paper, we propose a smart phase selection for a discrete IRS, using a trellis-based algorithm.   } The performance of the robust IRS-aided NOMA is compared with the non-robust and the perfect CSI scenarios in both cases of continuous and discrete IRS phase shifts. 
\textcolor{black}{We, then, establish a comparison benchmark by providing a solution for maximizing the sum-rate maximization in an IRS-aided OMA system.To this end, we present a robust IRS-aided OMA design using FDMA and TDMA. By comparing NOMA and OMA in three cases of perfect CSI, imperfect CSI robust design  and imperfect CSI non-robust design, we provide a new understanding of IRS-aided OMA and IRS-aided NOMA systems and their performance in different situations. }
% Moreover, it is shown that the IRS-assisted NOMA has {higher} spectral efficiency compared to its OMA counterparts. 
To better clarify, the contributions of this paper are summarized as follows:
\begin{itemize}
    \item In this paper, \textcolor{black}{for the first time,} channel estimation error is considered in the IRS-aided NOMA communication, and a robust design for IRS-assisted NOMA is presented.
    \item In order to design a joint beamforming technique for the IRS and the AP that is  robust to the channel estimation error,  the PDD technique is utilized. We first drive the rates of each user in the NOMA communication system. Then, using BSUM, we calculate a tractable lower bound for the sum-rate. 
    \item We maximize the lower bound of the sum-rate through a low complexity iterative algorithm, in each step of which the closed-form solutions for the optimization variables are calculated. 
    \item { We {also} consider the discrete phase shifts in the IRS, {rather than using the continuous ones}. We propose a smart phase selection {policy} based on the trellis algorithm which ultimately requires much less complexity compared to the exhaustive search method, and generates better results compared to the quantization approach. } 
    \item To further evaluate our proposed method, we address the same optimization problem in IRS-aided OMA communication systems, i.e.  IRS-FDMA and IRS-TDMA modes. 
    \item Based on our results we can claim that: i) the proposed method is fairly robust to channel estimation error, ii) the computational complexity of the proposed algorithm is very low compared to other methods in the literature, \textcolor{black}{iii) IRS-assisted NOMA outperforms IRS-assisted OMA in terms of spectral efficiency in perfect CSI scenario and imperfect CSI scenario enduring a certain level of channel uncertainty. }
\end{itemize}

% \subsection{Paper Organization}
The rest of this paper is organized as follows. In section \ref{System Model}, the  system model is presented. Section \ref{Section IRS-Aided NOMA} is dedicated to the problem formulation of IRS-aided NOMA and providing the PDD-based algorithm as a proper solution. Section \ref{Complexity Analysis} provides a complexity analysis for the IRS-aided NOMA design. In Section \ref{Section IRS-Aided OMA} the IRS-assisted OMA system is presented and the the sum-rate maximization problem is solved for  FDMA and  TDMA modes. Section \ref{Section Numerical Results}
includes the simulation results for the presented schemes, and finally Section \ref{Section Conclusion} concludes this paper.
% \subsection{Notations}

Throughout the paper, the variables, constants, vectors and matrices are represented by small italic letters, capital italic letters, small bold letters and capital bold letters, respectively. If $\mb{A}$ represents a matrix, the element in its $i$th row and $j$th column is represented by $a_{ij}$, and its $i$th column is referred to by $\mb{a}_i$. The notation $|.|$ denotes the absolute of a variable and the notation $||.||$ stands for the norm of a vector or the Frobenius norm of a matrix, depending on the argument. Also, $\re\{.\}$, $(.)^*$ and $(.)^H$ stand for the real part of a complex variable, the conjugate of a complex variable and the conjugate transpose of a complex vector/matrix, respectively.

\section{System Model }
\label{System Model}
Consider the downlink transmission of an AP with $N$ antennas to $K$ single-antenna users.  This transmission is aided by an IRS with $M$ phase shifters. A smart controller in the AP is responsible for managing the IRS by sharing information and coordinating the transmission. The power of the signals that are reflected by the IRS multiple times is much smaller than that of the signal reflected once, and thus it can be ignored \cite{Wu2019IntelligentBeamforming}. We  consider  an  indoor  application {undergoing rich scattering propagation environment};  hence,  we  consider  only  non-line-of-sight (NLOS) channels in our model \cite{YuEnablingSurfaces2019,HuangReconfigurable2018,AbdullahOptimizationNetworks2020,AbdullahAAnalysis2020,HouMIMOModel2019}.

\begin{figure}[t]
\begin{center}
  \includegraphics[scale=.52]{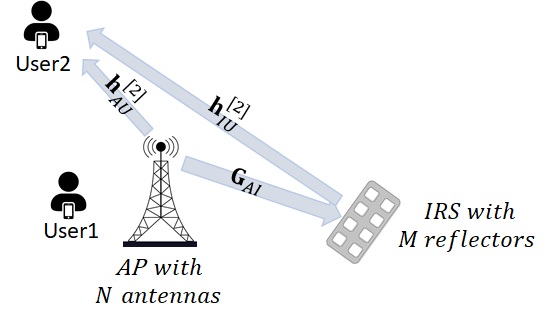}
      \caption{System Model with $K=2$ users.}
    \label{SysMod}
\end{center}
\end{figure}

The system model is depicted in Fig. \ref{SysMod}. According to this figure, the channel vectors from the AP to the $i$th user and from the IRS to this user are denoted by $\mb{h}_{AU}^{[i]}\in\mathbb{C}^{1\times N}$ and $\mb{h}_{IU}^{[i]}\in\mathbb{C}^{1\times M}$, respectively, and each element of  these vectors follows the complex normal distribution, i.e. $\mb{h}_{AU}^{[i]}\sim\mathcal{CN}\left(0,\beta_{AU}^{[i]}\mb{I}\right)$ and $\mb{h}_{IU}^{[i]}\sim\mathcal{CN}\left(0,\beta_{IU}^{[i]}\mb{I}\right)$. The channel matrix between the AP and the IRS is represented by $\mb{G}_{AI}^{}\in\mathbb{C}^{M\times N}$ which also follows the complex normal distribution as $\vect(\mb{G}_{AI})\sim\mathcal{CN}\left(0,\beta_{AI}\mb{I}\right)$. The notations $\beta_{AU}^{[i]}$, $\beta_{IU}^{[i]}$ and $\beta_{AI}$ represent the large-scale fading coefficients of their corresponding channels.

It is assumed that the imperfect cascaded AP-IRS-user channels are available at the AP and the source of this imperfection is the mobility of the users.  Hence, we have $\mb{h}_{AU}^{[i]}=\hat{\mb{h}}_{AU}^{[i]}+\tilde{\mb{h}}_{AU}^{[i]}$ and $\mb{h}_{IU}^{[i]}=\hat{\mb{h}}_{IU}^{[i]}+\tilde{\mb{h}}_{IU}^{[i]}$,
where $\hat{\mb{h}}_{AU}^{[i]}$ and $\hat{\mb{h}}_{IU}^{[i]}$ represent the channel estimation vectors whereas $\tilde{\mb{h}}_{AU}^{[i]}$ and $\tilde{\mb{h}}_{IU}^{[i]}$ stand for the channel estimation error vectors with the following distributions
\begin{equation}
    \tilde{\mb{h}}_{AU}^{[i]}\sim\mathcal{CN}\left(0,\sigma_{AU}^{2}\mb{I}\right),
\end{equation}
\begin{equation}
    \tilde{\mb{h}}_{IU}^{[i]}\sim\mathcal{CN}\left(0,\sigma_{IU}^{2}\mb{I}\right).
\end{equation}

\textcolor{black}{In this paper, we consider  continuous and discrete phase shifters  for the IRS:} 
\textcolor{black}{    \begin{itemize}
        \item $\mb{Continuous \ IRS}$: In an ideal case, the IRS phase shifters are continuous, i.e. the set of phase shifts can be expressed as $\mb{\Psi}_I=\{\theta_i|\theta_i=e^{j\psi_i},\psi_i\in[0,2\pi]\}$.
    \item $\mb{Discrete \ IRS}$: In a non-ideal scenario, the phase shifts are selected from a discrete set of phases, i.e. $\mb{\Psi}_N=\{\theta_i|\theta_i=e^{j\psi_i},\psi_i\in\mathcal{B}\}$, where $\mathcal{B}=\{e^{\frac{j2\pi m_j}{M_{IRS}}},m_j=1,...,M_{IRS}\}$ and $M_{IRS}$ denotes  the  number  of  possible  phases  that  can  be selected  by  each  phase  shifter  of  the  IRS. In other words, $M_{IRS}$ is the resolution of the phase shifters. 
    \end{itemize}}
In the following, the problem of sum-rate maximization is solved for IRS-NOMA for both cases of continuous and discrete IRS.

\section{IRS-Aided NOMA}\label{Section IRS-Aided NOMA}
\textcolor{black}{In this paper, we consider a  scenario where two users with different clusters are being served. In other words, we aim to investigate the effects of inter-cluster interference rather than intra-cluster effect. Clusters here  refer to classification of effective channels into superior channels and inferior channels in a NOMA system. Thus, throughout this paper, we assume $K=2$.  } 
With these assumptions, the information bearing vector $\mathbf{s} \in \mathbb{C}^{2 \times 1}$ is expressed as $\mb{s}=\left[\alpha_{1} s^{[1]}, \alpha_{2} s^{[2]}\right]^{T}$,
where $s^{[i]}$ represents the signal and $\alpha_i$ denotes the  corresponding power allocation coefficient for the $i$-th user satisfying  $\alpha_{1}^{2}+\alpha_{2}^{2}=1$.
Let $i$ and $k$ denote the index of  two users served by IRS-aided NOMA.  In this case, $\mb{x}^{[i]}\triangleq\mb{w}_{i}s^{[i]}$ denotes the transmit vector intended for the $i$-th user, where $\mb{w}_i$ is the $i$th active beamforming vector at the AP. The signal received by the $i$-th user is then written by
\begin{align}\label{received signal1}
y^{[i]}
% &=\sum_{j=1}^{2}\left[\mb{h}_{AU}^{[i]}\mb{w}_{j}\alpha_{j}s^{[j]}+\mb{h}_{IU}^{[i]}\mb{\Psi}\mb{G}_{AI}\mb{w}_{j}\alpha_{j}s^{[j]}\right]+n^{[i]}\nonumber\\
&=\left(\mb{h}_{AU}^{[i]}+\mb{h}_{IU}^{[i]}\mb{\Psi}\mb{G}_{AI}\right)\alpha_i\mb{x}^{[i]}+\left(\mb{h}_{AU}^{[i]}+\mb{h}_{IU}^{[i]}\mb{\Psi}\mb{G}_{AI}\right)\alpha_k\mb{x}^{[k]}+n^{[i]}\nonumber\\
&=\alpha_{i}\mb{h}^{[i]}\mb{x}^{[i]}+\alpha_{k}\mb{h}^{[i]}\mb{x}^{[k]}+n^{[i]}
\end{align}
where  $\mb{\Psi}$ represents the diagonal phase shift matrix of the IRS,  the effective AP-user channel vector $\mb{h}^{[i]}$ is defined by
$\mb{h}^{[i]}=\mb{h}_{AU}^{[i]}+\mb{h}_{IU}^{[i]}\mb{\Psi}\mb{G}_{AI}$ and $n^{[i]}$ is the additive white Gaussian noise with variance $\sigma_n^2$. Now, by reformulating the received signal in (\ref{received signal1}) in terms of the estimated channels, we have 
\begin{align}\label{received signal2}
    y^{[i]}&=\left(\hat{\mb{h}}_{AU}^{[i]}+\hat{\mb{h}}_{IU}^{[i]}\mb{\Psi}\mb{G}_{AI}\right)\alpha_{i}\mb{x}^{[i]}+\left(\hat{\mb{h}}_{AU}^{[i]}+\hat{\mb{h}}_{IU}^{[i]}\mb{\Psi}\mb{G}_{AI}\right)\alpha_{k}\mb{x}^{[k]}\nonumber\\
&\quad  +\sum_{j=1}^{2}\left[\left(\tilde{\mb{h}}_{AU}^{[i]}+\tilde{\mb{h}}_{IU}^{[i]}\mb{\Psi}\mb{G}_{AI}\right)\alpha_{j}\mb{x}^{[j]}\right]+n^{[i]}\nonumber\\
&=\alpha_{i}\hat{\mb{h}}^{[i]}\mb{x}^{[i]}+\alpha_{k}\hat{\mb{h}}^{[i]}\mb{x}^{[k]}+\sum_{j=1}^{2}\left[\alpha_{j}\tilde{\mb{h}}^{[i]}\mb{x}^{[j]}\right]+n^{[i]},
\end{align}
where 
\begin{equation}\label{compute g_hat}
    \hat{\mb{h}}^{[i]}\triangleq\hat{\mb{h}}_{AU}^{[i]}+\mb{v}\hat{\mb{H}}^{[i]}_c,
\end{equation}
\begin{equation}\label{compute g_tilda}
    \tilde{\mb{h}}^{[i]}\triangleq\tilde{\mb{h}}_{AU}^{[i]}+\mb{v}\tilde{\mb{H}}^{[i]}_c,
\end{equation}
\begin{equation}\label{compute v from psi}
\mb{v}\triangleq\left(\diag\left(\mb{\Psi}\right)\right)^T=\left[\theta_1,\dots,\theta_M\right].
\end{equation}
In (\ref{compute g_hat}) and (\ref{compute g_tilda}), $\hat{\mb{H}}^{[i]}_c=\diag\left(\hat{\mb{h}}_{IU}^{[i]}\right)\mb{G}_{AI}$ is the estimated cascaded AP-IRS-user channel and  $\tilde{\mb{H}}^{[i]}_c=\diag\left(\tilde{\mb{h}}_{IU}^{[i]}\right)\mb{G}_{AI}$ represents the cascaded channel estimation error. The distribution of the effective channel estimation error vector $\tilde{\mb{h}}^{[i]}$ for large values of $M$ is approximated by
\begin{align}\label{distribution of gtilda}
    \tilde{\mb{h}}^{[i]}\sim \mathcal{CN}\left(0,\sigma_{h}^2\mb{I}\right),
\end{align}
where $\sigma_{h}^2=\sigma_{AU}^2+M\sigma_{IU}^2\beta_{AI}$.
The approximation in (\ref{distribution of gtilda}) is derived in the appendix of \cite{Omiddd1} where it is shown that this approximation becomes more accurate when  number of IRS elements is large. \textcolor{black}{Also, it can be easily shown that this approximation remains valid whether the IRS phase shifts are continuous or discrete. }
Without loss of generality, we assume that the {effective} channel conditions of the first user is better than the second one \cite{LiuOnOMA2016}, i.e. $|\hat{\mb{h}}^{[1]}\mb{w}_{1}|^2\geq|\hat{\mb{h}}^{[2]}\mb{w}_{2}|^2$. With this assumption, the following theorem can be presented. 

\begin{theorem}
The achievable sum rate of the users is obtained by 
\begin{equation}
    R^{[sum]}=R^{[1]}+R^{[2]},
\end{equation}
in which 
\begin{equation}
    R^{[1]}=\log_2 \left(1+\frac{\alpha_{1}^{2}|\hat{\mb{h}}^{[1]}\mb{w}_{1}|^2}{\sigma_{h}^{2}\left(||\mb{w}_{1}||^2+||\mb{w}_{2}||^2\right)+\sigma_{n}^2}\right),
\end{equation}
is the minimum achievable rate of the first user and 
\begin{equation}
    R^{[2]}=\log_2 \left(1+\frac{\alpha_{2}^{2}|\hat{\mb{h}}^{[2]}\mb{w}_{2}|^2}{\alpha_{1}^{2}|\hat{\mb{h}}^{[2]}\mb{w}_{1}|^2+\sigma_{h}^{2}\left(||\mb{w}_{1}||^2+||\mb{w}_{2}||^2\right)+\sigma_{n}^2}\right),
\end{equation}
is the minimum achievable rate of the second user.
\end{theorem}
The proof of Theorem 1 is given in Appendix \ref{appendixB}.

In this paper, we aim to jointly optimize the active beamforming at the AP and the phase shifts at the IRS to maximize the sum achievable data rate. Specifically, the optimization problem is formulated as follows
\begin{equation}\label{main opt}
\begin{array}{cl}
\max\limits_{\mb{w}_1, \mb{w}_2, \mb{v}, \alpha_1, \alpha_2}& R^{[sum]}\\ 
\st 
%&\log\left(1+\frac{|\hat{\mb{h}}^{[i]}\mb{w}_{i}|^2}{\sum_{\substack{k=1\\k\neq i}}^{K}\left[|\hat{\mb{h}}^{[i]}\mb{w}_{k}|^2\right]+\sigma_{h}^{2}\sum_{k=1}^{K}\left[|\mb{w}_{k}|^2\right]+\sigma_{n}^2}\right)\geq \lambda_{i}, \forall i\\
%&\mb{\Psi}\mb{\Psi}^H=\mb{I},\\
&|\theta_i|=1, i=1,...,M,\\
&\tr \left(\mb{W}\mb{W}^H\right)\leq P, \\
&\alpha_{1}^{2}+\alpha_{2}^{2}=1,\\
&\left|\hat{\mb{h}}^{[1]}\mb{w}_{1}\right|^2\geq\left|\hat{\mb{h}}^{[2]}\mb{w}_{2}\right|^2.
%&\left|\mb{\Psi}_{jj}\right|=1, \forall j.
\end{array}
\end{equation}
where $\mb{W}=[\mb{w}_1,\mb{w}_2]$. It can be easily observed that this optimization problem is non-convex, with coupled variables.
\textcolor{black}{Specifically, the optimization problem (12) incorporates $R^{[1]}$ and $R^{[2]}$ which are nonlinear components with respect to $\mb{v}$, $\mb{W}_{i}$ and $\alpha_{i}$, and make the problem non-convex. Additionally, the forth constraint in (12) implies a strong and mutual coupling between $\mb{v}$ and $\mb{w}_{i}$ which can be hardly dealt with in its initial construction.}
To address these issues, in the following a low-complexity algorithm is presented for addressing this problem.  
\subsection{PDD-based Solution: Continuous Phase shifters}
In this section, we propose the PDD-based algorithm \cite{Pddmain1,PDDmain} to solve the problem in (\ref{main opt}). To this end, we first introduce a set of auxiliary variables as $\mathcal{S}=\{\mb{T},\mb{v},\mb{W},\mb{a},\bar{\mb{W}},\bar{\mb{T}},\bar{\mb{a}}\}$, where $\mb{T}=\mb{W}^H\hat{\mb{H}}^H$,  $\hat{\mb{H}}\triangleq\left[\hat{\mb{h}}_{}^{[1] \ T},\hat{\mb{h}}_{}^{[2] \ T}\right]^T$,  $\mb{a}=[\alpha_1,\alpha_2]^T$, $\bar{\mb{W}}=\mb{W}$, $\bar{\mb{T}}=\mb{T}$, and $\bar{\mb{a}}=\mb{a}$. With these definitions, the optimization problem becomes
\begin{equation}\label{second main opt}
\begin{array}{cl}
\max\limits_{\mathcal{S}}& \log_2 \left(1+\frac{\alpha_{1}^{2}|t_{11}|^2}{\sigma_{h}^{2}\left(||\mb{w}_{1}||^2+||\mb{w}_{2}||^2\right)+\sigma_{n}^2}\right)+\log_2 \left(1+\frac{\alpha_{2}^{2}|t_{22}|^2}{\alpha_{1}^{2}|t_{12}|^2+\sigma_{h}^{2}\left(||\mb{w}_{1}||^2+||\mb{w}_{2}||^2\right)+\sigma_{n}^2}\right)\\ 
\st 
&|\theta_i|=1, i=1,...,M,\\
&\tr \left(\bar{\mb{W}}\bar{\mb{W}}^H\right)\leq P, \\
&\bar{\mb{a}}^T\bar{\mb{a}}=1,\\
&\left|\bar{t}_{11}\right|^2\geq\left|\bar{t}_{22}\right|^2,\\
%&\left|\mb{\Psi}_{jj}\right|=1, \forall j.
&\mb{T}=\mb{W}^H\hat{\mb{h}}^H,
\bar{\mb{T}}=\mb{T}, \bar{\mb{W}}=\mb{W},
\bar{\mb{a}}=\mb{a}.
\end{array}
\end{equation}
Furthermore, to obtain the augmented Lagrangian problem, the equality conditions should be put into the objective function. This leads to the following optimization problem
\begin{equation}\label{Third main opt}
\begin{array}{cl}
\max\limits_{\mathcal{S}}& \log_2 \left(1+\frac{\alpha_{1}^{2}|t_{11}|^2}{\sigma_{h}^{2}\left(||\mb{w}_{1}||^2+||\mb{w}_{2}||^2\right)+\sigma_{n}^2}\right)+\log_2 \left(1+\frac{\alpha_{2}^{2}|t_{22}|^2}{\alpha_{1}^{2}|t_{12}|^2+\sigma_{h}^{2}\left(||\mb{w}_{1}||^2+||\mb{w}_{2}||^2\right)+\sigma_{n}^2}\right)-Q_{\gamma}(\mathcal{S})\\
\st 
&|\theta_i|=1, i=1,...,M,\\
&\tr \left(\bar{\mb{W}}\bar{\mb{W}}^H\right)\leq P, \\
&\bar{\mb{a}}^T\bar{\mb{a}}=1,\\
&\left|\bar{t}_{11}\right|^2\geq\left|\bar{t}_{22}\right|^2,\\
%&\left|\mb{\Psi}_{jj}\right|=1, \forall j.
%&\mb{T}=\mb{W}^H\hat{\mb{h}}^H,\\
%&\mb{T}=\bar{\mb{T}}, \mb{W}=\bar{\mb{W}},
%\mb{a}=\bar{\mb{a}}.
\end{array}
\end{equation}
where $Q_{\gamma}(\mathcal{S})$ is calculated by 
\begin{align}\label{P_rho}
\small
    Q_{\gamma}(\mathcal{S})=\frac{1}{2\gamma}\Big(&
    \sum_{i=1}^2||\mb{w}_i-\bar{\mb{W}}_i+\gamma \boldsymbol{\lambda}_{w_{i}}||^2+\sum_{i=1}^2\sum_{j=1}^2|t_{ij}-\mb{w}_i^H\hat{\mb{h}}^{[j]H}+\gamma {\lambda}_{{h}_{ij}}|^2\nonumber\\&
   +\sum_{i=1}^2\sum_{j=1}^2|{t}_{ij}-\bar{{t}}_{ij}+\gamma {\lambda}_{t_{ij}}|^2+\sum_{i=1}^2|\alpha_i-\bar{\alpha}_i+\gamma {\lambda}_{a_{i}}|^2
    \Big).
\end{align}
The penalty parameter for the Lagrangian function is denoted by $\gamma$, and $\boldsymbol{\lambda}_{w_i}$, $\lambda_{h_{ij}}$, $\lambda_{t_{ij}}$ and $\lambda_{a_i}$, $i,j\in\{1,2\}$, are the dual variables of the equality conditions in problem (\ref{second main opt}). The vector $\boldsymbol{\lambda}_{w_i}$ is the $i$th column of the matrix  $\boldsymbol{\Lambda}_w$ and $\lambda_{h_{ij}}$, $\lambda_{t_{ij}}$ and $\lambda_{a_i}$ are the elements of $\boldsymbol{\Lambda}_h$, $\boldsymbol{\Lambda}_t$ and $\boldsymbol{\lambda}_a$, respectively.
Note that, here one of the equality conditions is not put into the objective function. The reason is that by keeping this condition in the constrains, a closed-form solution can be calculated with projection. The details are explained later in this paper. 

Now to apply BSUM to (\ref{Third main opt}), we introduce the following theorem.
\begin{theorem} The objective function of the optimization problem in (\ref{second main opt}) can be rewritten as
\begin{align}\label{lemma1 equation}
    % &\log_2 \left(1+\frac{\alpha_{1}^{2}|t_{11}|^2}{\sigma_{h}^{2}\left(||\mb{w}_{1}||^2+||\mb{w}_{2}||^2\right)+\sigma_{n}^2}\right)\nonumber\\&\quad+\log_2 \left(1+\frac{\alpha_{2}^{2}|t_{22}|^2}{\alpha_{1}^{2}|t_{12}|^2+\sigma_{h}^{2}\left(||\mb{w}_{1}||^2+||\mb{w}_{2}||^2\right)+\sigma_{n}^2}\right)\nonumber\\&=
    \max\limits_{q_1,q_2,d_1,d_2} \sum_{i=1}^2\Big(\log(d_i)-d_if_i(q_i,\mathcal{S})\Big)+2,
\end{align}
where
\begin{align}
    f_1(q_1,\mathcal{S})&=|1-q_1^*\alpha_1t_{11}|^2+\sigma_{h}^2(||q_1^*\mb{w}_{1}||^2+||q_1^*\mb{w}_{2}||^2)+\sigma_n^2|q_1|^2,
\end{align}
\begin{align}
    f_2(q_2,\mathcal{S})&=|1-q_2^*\alpha_2t_{22}|^2+\alpha_1^2|q_2^*t_{12}|^2+\sigma_{h}^2(||q_2^*\mb{w}_{1}||^2+||q_2^*\mb{w}_{2}||^2)+\sigma_n^2|q_2|^2.
\end{align}
\end{theorem}
The proof of this theorem can be simply finished by using the first order equality condition on the equation (\ref{lemma1 equation}), to obtain the optimal values for $q_1$, $q_2$, $d_1$ and $d_2$ as
\begin{equation}
    q_1^{opt}(\mathcal{S})=\frac{\alpha_{1}t_{11}}{\alpha_{1}^{2}|t_{11}|^2+\sigma_{h}^{2}\left(||\mb{w}_{1}||^2+||\mb{w}_{2}||^2\right)+\sigma_{n}^2},\label{u1Opt}
\end{equation}
\begin{equation}
    q_2^{opt}(\mathcal{S})=\frac{\alpha_{2}t_{22}}{\alpha_2^2|t_{22}|^2+\alpha_{1}^{2}|t_{12}|^2+\sigma_{h}^{2}\left(||\mb{w}_{1}||^2+||\mb{w}_{2}||^2\right)+\sigma_{n}^2},\label{u2Opt}
\end{equation}
\begin{equation}
    d_1^{opt}(\mathcal{S})= 1+\frac{\alpha_{1}^{2}|t_{11}|^2}{\sigma_{h}^{2}\left(||\mb{w}_{1}||^2+||\mb{w}_{2}||^2\right)+\sigma_{n}^2},\label{w1Opt}
\end{equation}
\begin{equation}
    d_2^{opt}(\mathcal{S})=1+\frac{\alpha_{2}^{2}|t_{22}|^2}{\alpha_{1}^{2}|t_{12}|^2+\sigma_{h}^{2}\left(||\mb{w}_{1}||^2+||\mb{w}_{2}||^2\right)+\sigma_{n}^2}.\label{w2Opt}
\end{equation}
This theorem  gives a lower bound for the objective function of (\ref{second main opt}) as
\begin{align}\label{Lower Bound}
    &\log_2 \left(1+\frac{\alpha_{1}^{2}|t_{11}|^2}{\sigma_{h}^{2}\left(||\mb{w}_{1}||^2+||\mb{w}_{2}||^2\right)+\sigma_{n}^2}\right)+\log_2 \left(1+\frac{\alpha_{2}^{2}|t_{22}|^2}{\alpha_{1}^{2}|t_{12}|^2+\sigma_{h}^{2}\left(||\mb{w}_{1}||^2+||\mb{w}_{2}||^2\right)+\sigma_{n}^2}\right)\nonumber\\&\geq\sum_{i=1}^2\Big(\log(d_i^{opt}(\mathcal{S}))-d_i^{opt}(\mathcal{S})f_i(q_i^{opt}(\mathcal{S}),\mathcal{S})\Big)+2.
\end{align}
This locally tight lower bound is  tractable and by using it instead of its original equation, the BSUM can be applied to our problem. Then, the new form of the optimization problem becomes
\begin{equation}\label{second optimization 26}
\begin{array}{cl}
\min\limits_{\mathcal{S}}& \sum_{i=1}^2d_i(\mathcal{S})f_i(q_i(\mathcal{S}),\mathcal{S})+Q_{\gamma}(\mathcal{S}) \\ 
\st 
%&\log\left(1+\frac{|\hat{\mb{h}}^{[i]}\mb{w}_{i}|^2}{\sum_{\substack{k=1\\k\neq i}}^{K}\left[|\hat{\mb{h}}^{[i]}\mb{w}_{k}|^2\right]+\sigma_{h}^{2}\sum_{k=1}^{K}\left[|\mb{w}_{k}|^2\right]+\sigma_{n}^2}\right)\geq \lambda_{i}, \forall i\\
%&\mb{\Psi}\mb{\Psi}^H=\mb{I},\\
&|\theta_i|=1, i=1,...,M,\\
&\tr \left(\bar{\mb{W}}\bar{\mb{W}}^H\right)\leq P, \\
&\bar{\mb{a}}^T\bar{\mb{a}}=1,\\
&\left|\bar{t}_{11}\right|^2\geq\left|\bar{t}_{22}\right|^2.\\
%&\left|\mb{\Psi}_{jj}\right|=1, \forall j.
%&\mb{T}=\mb{W}^H\hat{\mb{h}}^H,\\
%&\bar{\mb{T}}=\mb{T}, %\bar{\mb{W}}=\mb{W},
\end{array}
\end{equation}

Now, by applying the BSUM, the optimization problem in (\ref{second optimization 26}) can be solved via the following steps, each resulting in a closed-form solution for a sub-set of the variables. 
% \textcolor{black}{In the following, the problem is solved for two cases of ideal and non-ideal IRS phase shifters. }
 
% \begin{enumerate}
    % \item 
    At first, we solve the problem based on $\mb{W}$ while considering all other variables as constants. Based on the first order optimality condition we can obtain the solution as
    \begin{align}\label{V1}
        \mb{w}_1=&\left(2\gamma|q_1|^2d_1\sigma_{h}^2\mb{I}+2\gamma|q_2|^2d_2\sigma_{h}^2\mb{I}+\mb{I}+\sum_{j=1}^2\hat{\mb{h}}^{[j]H}\hat{\mb{h}}^{[j]} \right)^{-1}\nonumber\\&\times\left(\bar{\mb{W}}_1-\gamma\boldsymbol{\lambda}_{w_1}+\sum_{j=1}^2(\hat{\mb{h}}^{[j]H}t_{1j}^*+\hat{\mb{h}}^{[j]H} \gamma \lambda_{h_{1j}}^* ) \right),
    \end{align}
    \begin{align}\label{V2}
        \mb{w}_2&=\left(2\gamma|q_1|^2d_1\sigma_{h}^2\mb{I}+2\gamma|q_2|^2d_2\sigma_{h}^2\mb{I}+\mb{I}+\sum_{j=1}^2\hat{\mb{h}}^{[j]H}\hat{\mb{h}}^{[j]}  \right)^{-1}\nonumber\\&\times\left(\bar{\mb{W}}_2-\gamma\boldsymbol{\lambda}_{w_2}+\sum_{j=1}^2(\hat{\mb{h}}^{[j]H}t_{2j}^*+\hat{\mb{h}}^{[j]H} \gamma \lambda_{h_{2j}}^* ) \right).
    \end{align}
    Secondly, we find the optimal values for $\alpha_1$ and $\alpha_2$ based on the first order optimality condition, as
    \begin{align}
        &\alpha_1=\frac{2\gamma d_1 \re\{q_1 t_{11}^*\}+\bar{\alpha}_1-\gamma \lambda_{a_1}}{1+2\gamma d_1|q_1|^2|t_{11}|^2+2\gamma d_2|q_2^*t_{12}|^2},\label{a1}
        \\&\alpha_2=\frac{2\gamma d_2 \re\{q_2^* t_{22}\}+\bar{\alpha}_2-\gamma \lambda_{a_2}}{1+2\gamma d_2|q_2|^2|t_{22}|^2}.\label{a2}
    \end{align} 
    In the next step, we use the same method for the variable $\mb{T}$. The closed-form solution for each element of this variable is
    \begin{equation}
        t_{11}=\frac{2\gamma d_1q_1\alpha_1+\mb{w}_1^H \hat{\mb{h}}^{[1]H}+\bar{t}_{11}-\gamma(\lambda_{h_{11}}+\lambda_{t_{11}}) }{2\gamma d_1|q_1|^2\alpha_1^2+2},\label{x11}
    \end{equation}
    \begin{equation}
        t_{12}=\frac{\mb{w}_1^H \hat{\mb{h}}^{[2]H}+\bar{t}_{12}-\gamma(\lambda_{h_{12}}+\lambda_{t_{12}}) }{2\gamma d_2\alpha_1^2|q_2|^2+2},\label{x12}
    \end{equation}
    \begin{equation}
        t_{21}=\frac{1}{2}\left(\mb{w}_2^H \hat{\mb{h}}^{[1]H}+\bar{t}_{21}-\gamma(\lambda_{h_{21}}+\lambda_{t_{21}}) \right),\label{x21}
    \end{equation}
    \begin{equation}
        t_{22}=\frac{2\gamma d_2q_2\alpha_2+\mb{w}_2^H \hat{\mb{h}}^{[2]H}+\bar{t}_{22}-\gamma(\lambda_{h_{22}}+\lambda_{t_{22}}) }{2\gamma d_2|q_2|^2\alpha_2^2+2}.\label{x22}
    \end{equation}
Then, the optimization problem with respect to $\bar{\mb{T}}$ is solved, as 
\begin{equation} \label{solutionX}
    \bar{\mb{T}}=\begin{cases}
    \mb{Y}, & |y_{11}|^2\geq |y_{22}|^2\\
    \begin{bmatrix}
    \frac{y_{11}-y_{22}}{2}&  y_{12} \\y_{21}& \frac{y_{11}-y_{22}}{2}
    \end{bmatrix} \text{or} 
    \begin{bmatrix}
    \frac{y_{11}+y_{22}}{2}&  y_{12} \\y_{21}& \frac{y_{11}+y_{22}}{2} 
    \end{bmatrix}, & |y_{11}|^2< |y_{22}|^2
    \end{cases}.
\end{equation}
In (\ref{solutionX}), $\mb{Y}=\mb{T}+\gamma\boldsymbol{\Lambda}_{x}$ and $y_{ij}$ represents the element in the  $i$th row and the $j$th column of  matrix $\mb{Y}$. Also, in case $|y_{11}|^2< |y_{22}|^2$, the matrix with the least Euclidean distance to $\mb{Y}$ would be selected as $\bar{\mb{T}}$.

    % \item 
    Next, we need to consider the auxiliary variables $\bar{\mb{W}}$ and $\bar{\mb{a}}$. The corresponding constraint for $\bar{\mb{W}}$ is a projection of a point onto a ball centered at the origin\footnote{ The projection of a point $\mb{T}$ into a set $\mathcal{X}$ is defined by
     $\min\limits_{\mb{P} \in \mathcal{X}}\|\mb{T}-\mb{P}\|$.
In case $\mathcal{X}$ is a sphere centered at the origin with radius $R$ ($\mathcal{X}=\{\mb{T} |\|\mb{T}\| \leq R\}),$ the projection of $\mb{T}$ is obtained by $R\frac{\mb{T}}{\|\mb{T}\|+\max (0, R-\|\mb{T}\|)}$.}. The closed-form solution for this exists as
    \begin{equation}\label{V_Bar}
        \bar{\mb{W}}=\mathcal{P}_{P}(\mb{W}+\gamma\boldsymbol{\Lambda}_w).
    \end{equation}
    Similarly, the corresponding constraint for $\bar{\mb{a}}$ can be satisfied by projection of a point onto a circle and the closed-form solutions for  $\bar{\alpha}_1$ and $\bar{\alpha}_2$ are calculated as
    \begin{align}
        &\bar{\alpha}_1=\sqrt{\frac{1}{1+(\frac{\alpha_2+\gamma \lambda_{a_2}}{\alpha_1+\gamma \lambda_{a_1}})^2}},\label{a_BAR1}
        \\&\bar{\alpha}_2=(\frac{\alpha_2+\gamma \lambda_{a_2}}{\alpha_1+\gamma \lambda_{a_1}})\bar{\alpha}_1.\label{a_BAR2}
    \end{align}
    % \item 
    \quad The final step is to solve the problem for the variable $\mb{v}$. In this case, the problem could be rewritten as 
    \begin{equation}\label{ProblemF}
    \begin{array}{cl}
    \min\limits_{|\theta_i|=1,i=1,...,M}& \mb{v}\mb{A}\mb{v}^H-2\re\{\mb{c}\mb{v}^H\},
    \end{array}
    \end{equation}
    where,
    \begin{equation} 
        \mb{A}=\frac{1}{2\gamma}\sum_{i=1}^K\sum_{j=1}^K\hat{\mb{H}}^{[j]}_c\mb{w}_i\mb{w}_i^H\hat{\mb{H}}^{[j]H}_c,\label{40}
    \end{equation}
    \begin{equation}  
    \small
        \mb{c}=\frac{1}{2\gamma}\sum_{i=1}^K\sum_{j=1}^K\left(t_{ij}^*+\gamma \lambda_{h_{ij}}^*-\hat{\mb{h}}_{AU}^{[j]}\mb{w}_i \right)\mb{w}_i^H\hat{\mb{H}}^{[j]H}_c.\label{41}
    \end{equation}
    Now for one specified $\theta_k, k=1,...,M$, the optimization problem becomes 
    \begin{equation}
    \begin{array}{cl}
    \min\limits_{|\theta_k|=1}& |\theta_k|^2{a}_{kk}-2\re\{({c}_k-\sum_{i\neq k}^{M}\theta_i{a}_{ik})\theta_k^*\},
    \end{array}
    \end{equation}
    and the solution to this optimization problem is 
    \begin{equation}
    \theta_k=\frac{{c}_k-\sum_{i\neq k}^{M}\theta_i{a}_{ik}}{|{c}_k-\sum_{i\neq k}^{M}\theta_i{a}_{ik}|}.
    \end{equation}
    Then, an iterative algorithm is proposed to alternately optimize one phase shift while keeping the others fixed until final convergence.

The steps of  the PDD algorithm using BSUM to solve the optimization problem in (\ref{second optimization 26}) are described in Algorithm \ref{Alg22}. \textcolor{black}{The steps of the basic PDD method in which the dual variables and the penalty parameter are updated, in addition to a detailed convergence analysis for this scheme are provided in \cite{Pddmain1}. Specifically, it is shown that under appropriate conditions, the sequence of $\mb{t}^{k}\in\mathcal{S}$ generated by the PDD method tends to a KKT  point of the main problem.} In Algorithm 1, the update method for the dual variable $\boldsymbol{\lambda}_a$ in the $k$th iteration is as follows
\begin{equation}\label{update dual Lagrangian}
    \boldsymbol{\lambda}_a^{k}=\boldsymbol{\lambda}_{a}^{k-1}+\frac{1}{\gamma}(\mb{a}-\bar{\mb{a}}).
\end{equation}
The other dual Lagrangian variables, e.g. $ \boldsymbol{\Lambda}_h, \boldsymbol{\Lambda}_w, \boldsymbol{\Lambda}_t$ are updated in the same manner. Also, the penalty parameter in the $k$th iteration is updated by
\begin{equation}\label{update rho}
    \gamma^{k}=\zeta\gamma^{k-1},
\end{equation}
where $\zeta \in (0,1)$ is a decreasing parameter.

\begin{algorithm} 
\caption{The PDD algorithm using BSUM to solve the problem in  (\ref{second optimization 26})}
\begin{algorithmic} \label{Alg22}
\STATE $\mb{Initialize }$ \ $\gamma, \boldsymbol{\Lambda}_h, \boldsymbol{\Lambda}_w, \boldsymbol{\Lambda}_t, \boldsymbol{\lambda}_a, \epsilon, \eta, \zeta$
\STATE $\mb{Initialize }$  \ $\mb{v},\mb{W}$ and $\mb{a}$ such that all constraints are met
\STATE $\mb{Compute}$ $\hat{\mb{h}}$ based on (\ref{compute g_hat}) 
\STATE $\mb{Set}$ \quad \quad \quad  $\mb{T}=\mb{W}^H\hat{\mb{h}}^H, \bar{\mb{W}}=\mb{W}$, $\bar{\mb{T}}=\mb{T}$, $\bar{\mb{a}}=\mb{a}$
\STATE $\mb{Repeat}$
\STATE \quad 1. Compute $q_i^{opt}$ and $d_i^{opt}$  by (\ref{u1Opt})-(\ref{w2Opt})
% \STATE \quad 2. compute $\mb{B}$, $\mb{A}$, $\mb{D}$ and $b$ by (\ref{mb{B}}), (\ref{mb{A}}), (\ref{mb{D}}) and (\ref{b})
\STATE \quad 2. Compute $\mb{W}$ by (\ref{V1}), (\ref{V2})
\STATE \quad 3. Compute $\mb{a}$ by (\ref{a1}), (\ref{a2})
\STATE \quad 4. Compute $\mb{T}$ by (\ref{x11})-(\ref{x22})
\STATE \quad 5. Compute $\bar{\mb{T}}$ by (\ref{solutionX})
\STATE \quad 6. Compute $\bar{\mb{W}}$, $\bar{\mb{a}}$ by (\ref{V_Bar}), (\ref{a_BAR1}), (\ref{a_BAR2})
\STATE \quad 7. Compute $\mb{v}$ by solving (\ref{ProblemF}) in an iterative algorithm
\STATE \quad 8. Recalculate $\mb{\hat{G}}$ with the updated $\mb{v}$ by (\ref{compute g_hat}) 
\STATE \quad 9. Calculate the Frobenius norms of $||\mb{T}-\mb{W}^H\hat{\mb{h}}^H||$,\\ \quad \quad $||\mb{W}-\bar{\mb{W}}||$, $||\mb{T}-\bar{\mb{T}}||$ and $||\mb{a}-\bar{\mb{a}}||$.
\STATE \quad 10. If\ the\ value\ of\ calculated\  norms\ $\leq \eta $
\STATE \quad \quad \quad Update $\boldsymbol{\Lambda}_h, \boldsymbol{\Lambda}_w, \boldsymbol{\Lambda}_t, \boldsymbol{\lambda}_a$ as in (\ref{update dual Lagrangian})
\STATE \quad \quad else
\STATE \quad \quad \quad Update $\gamma$ by (\ref{update rho})
\STATE \quad \quad end\ if
\STATE $\mb{Until}$ the absolute difference of two consecutive sum-rates is less than $\epsilon$.
\end{algorithmic}
\end{algorithm}

\subsection{PDD-based Solution: Discrete Phase Shifters}
\textcolor{black}{In case of  discrete phase shifters for the IRS, the phase of each IRS element should be optimally chosen from a set of $M_{IRS}$ discrete phase shifts. In this case, to solve the problem in (\ref{second optimization 26}), we employ the PDD algorithm as in the previous sub-section. All of the  steps of the algorithm are the same as the continuous IRS case, except for the seventh step in which the variable $\mb{v}$ is calculated. }
\textcolor{black}{In this case, the problem in (\ref{ProblemF}) can be rewritten as
    \begin{equation}\label{ProblemF discrete}
    \begin{array}{cl}
    \min\limits_{\theta_k\in\mb{\Psi}_N,k=1,...,M}& 
    R_{trellis},
    \end{array}
    \end{equation}
    where 
    \begin{equation}\label{trellis benchmark}
         R_{trellis}=\sum_{j=1}^M\re\left\{(\sum_{i=1}^{j-1}\theta_j^*a_{ij}\theta_i)-c_j\theta_j^*\right\}.
    \end{equation}
    Note that the objective function in (\ref{ProblemF discrete}) is in the form of a consecutive real summation; therefore we can use the trellis algorithm to find the best discrete IRS phase shifts \cite{Kazemi2017}. }
    % \textcolor{\inquirycolor}{(Trellis is a general trend. You can use your IRS-trellis Arxiv as the reference. Also you can add the proof of this objective function to the appendices)}}
    
\begin{figure}[t]  \label{Trellis-based design for psi}
\begin{center}
   \includegraphics[scale=.94]{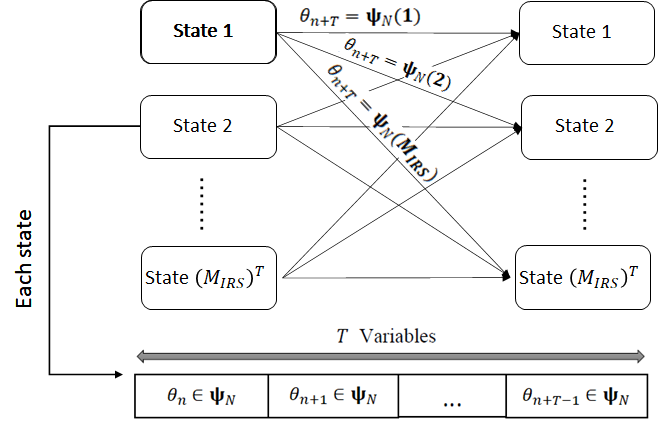}
       \caption{Trellis-based design for determination of $\mb{v}$.}
    \label{fig:my_label}
\end{center}
\end{figure}

\textcolor{black}{For the sake of solving  Problem (\ref{ProblemF discrete}), let $T$ be the number of variables ($\theta_i$) as the memory, each having $M_{IRS}$ possible choices. Therefore, as depicted in Fig. \ref{fig:my_label}, we construct a trellis with $(M_{IRS})^{T}$ states, each having $M_{IRS}$ outgoing branches with labels selected from $\mb{\Psi}_{N}$.}

\textcolor{black}{We describe the details of the proposed algorithm as follows. At  first, the initial $T$ variables are taken as initial memory values and their $(M_{IRS})^T$ possible permutations form the trellis states. Initial benchmarks are calculated by inserting the values of each state in the first $T$ terms of the objective function of (\ref{ProblemF discrete}). At the $n$-th stage, the branch labels imply the $(n+T)$-th variable ($\theta_{n+T}$), and the branch benchmark is the ($n+T$)-th term of the objective function of (\ref{ProblemF discrete}). At each stage, the cumulative benchmark of branches are calculated by adding the branch benchmarks to the cumulative benchmark of their originating paths. After that, among the branches entering the same state, the branch with the least cumulative benchmark is kept and the others are removed. The algorithm is terminated after $M-T$ stages and the path with the minimum cumulative benchmark is selected as the final solution. The labels on the selected path represent the latter $M-T$ variables, and the initial state associated with the selected path stands for the first $T$ variables. This process is shown in Algorithm \ref{Alg1}. The overall solution is the same as Algorithm \ref{Alg22}, but instead of its 7th step, we employ the method in Algorithm \ref{Alg1}.  }  For more details on the trellis algorithm please refer to \cite{Omid2022Trellis}. 
\begin{algorithm} 
\caption{Trellis-based discrete phase shift determination for the IRS}
\textcolor{black}{\begin{algorithmic}\label{Alg1}
\STATE $\mb{Input}:$ $\mb{\Psi}_{N},\mb{v}^{Initial}$
\STATE $\mb{Output}:$ $\mb{v}$
\STATE $\mb{Initialize}$ all possible permutations of $\theta_1,\dots,\theta_T$
\FOR{$i=T+1,\dots,M$}
\FOR{$j=1,\dots,M_{IRS}$}
\STATE $\theta_i=\mb{\Psi}_{N}(j)$;
\STATE calculate (\ref{trellis benchmark}) as the benchmark;
\ENDFOR
\STATE eliminate all paths except the one with the minimum benchmark value;
\ENDFOR
\STATE Choose $\theta_i,\dots,\theta_{M}$, such that they lead to the minimum cumulative benchmark value.
\end{algorithmic}}
\end{algorithm}

\section{Complexity Analysis}\label{Complexity Analysis}
In this section, the complexity of the proposed method is calculated for both cases of continuous and discrete IRS, and it is compared with other techniques in the literature.
\subsection{Continuous IRS}
The proposed scheme in Algorithm \ref{Alg22} is iterative. In each iteration, the computational complexity consists of calculating the variables in steps $1$ through $10$ of Algorithm \ref{Alg22} and updating either the dual Lagrangian variables or the penalty parameter. Note that in step $10$, the order of complexity is highest when the dual variables are being updated. Also, for step 7 of the continuous IRS case, the order of computational complexity is approximated by the computational complexity of equations  (\ref{40}) and (\ref{41}) since it relies on $M$ which is large. Considering a $K=2$ user system, the order of complexity of each step in Algorithm \ref{Alg22}, for a continuous IRS, is
\begin{equation}\label{com1}
    \mathcal{O}(6M^2+(2N^2+8N)M+2N^3+21N^2+157N).
\end{equation}

\begin{figure}[t]
\begin{center}
  \includegraphics[scale=.35]{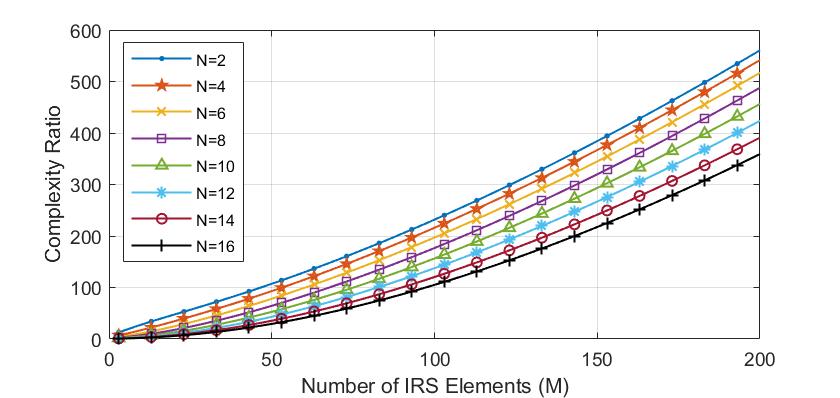}
      \caption{Complexity ratio each iteration of the algorithm in \cite{MuExploitingOptimization2020} to each iteration of our PDD-based proposed algorithm.}
    \label{fig:complexity}
\end{center}
\end{figure}

% Table \ref{complexity table 1} illustrates the order of complexity for each step of Algorithm \ref{Alg22}, assuming there are only $K=2$ users in the system. Regarding Table \ref{complexity table 1}, note that for step 7, the order of computational complexity is approximated by the computational complexity of equations  (\ref{40}) and (\ref{41}) since it relies on $M$ which is large. %since the rest of the algorithm to find the optimal $\mb{v}$ has very low complexity and according to our simulations, it converges in usually 2-3 steps. 
% Finally, the overall order of complexity for each step in  of Algorithm \ref{Alg22} is
% $\mathcal{O}(6M^2+(2N^2+8N+9)M+2N^3+21N^2+157N+440)$. 
This order of complexity is lower than that of the methods in the literature focusing on sum-rate maximization in the downlink.
% In \cite{ZengSumNOMA2021}, the authors used an semi-definite relaxation (SDR)-based iterative algorithm to find near-optimal solutions for the sum-rate maximization problem in IRS-aided NOMA. 
In \cite{MuExploitingOptimization2020}, an alternating optimization technique in used to find optimal active AP beamforming vector and the passive IRS phase shifts for the sum-rate maximization problem. The authors used the interior-point solver in CVX to solve the resulting convex optimization  problems in each iteration of their algorithm which scales up the order of complexity.  For instance the complexity of each iteration of the algorithm proposed in \cite{MuExploitingOptimization2020} is 
\begin{equation}\label{com2}
    \mathcal{O}(\max(N,3K(K-1)^4)\sqrt{N}\log(\frac{1}{\mu_c})+(3K^2+M)^{3.5}),
\end{equation}
 where $\mu_c$ is the accuracy defined in \cite{MuExploitingOptimization2020}. Specifically speaking, consider a simple case where $N=K=2$ and $M=50$, which are low dimensions for the system model, and assume that $\mu_c=0.1$. In this scenario, the order of complexity of the algorithm proposed in \cite{MuExploitingOptimization2020}  is 100 times the complexity of our proposed scheme.
Fig. \ref{fig:complexity} better demonstrates the complexity ratio of the algorithm in \cite{MuExploitingOptimization2020} to our PDD-based proposed method in different system dimensions. As shown in this figure and based on the equations (\ref{com1}) and (\ref{com2}), the complexity of our method rises with $N^3$ while the complexity of the other algorithm rises with $N$, but still even for large values of $N$, the complexity of our method is far less.

\subsection{Discrete IRS} \label{Discrete IRS}
\textcolor{black}{In case of the discrete IRS, the 7th step of the Algorithm \ref{Alg22} is replaced with Algorithm \ref{Alg1}. The proposed  trellis-based algorithms consist of  $M-T$ stages. The optimization of $\mb{v}$ consists of $(M_{IRS})^{T}$ states with $M_{IRS}$ branches entering each state. Hence, a total of $(M-T)(M_{IRS})^{T+1}$ comparisons are needed. This number is negligible compared with an exhaustive search method with $(M_{IRS})^{M}$ comparisons, especially for large values of $M$. By using the trellis-based algorithm, the computational complexity of each step of Algorithm \ref{Alg22} becomes
\begin{equation}\label{comDis}
    \mathcal{O}(2M^2+4NM+2N^3+20N^2+150N+(M-T)(M_{IRS})^{T+1}).
\end{equation}
In case of using exhaustive search instead of trellis, the last term on (\ref{comDis}) is replaced by $(M_{IRS})^{M}$, while if we use the quantization technique the last term of (\ref{comDis}) is replaced by $M$. To better understand, consider a system model with the following variables $K=2$, $M=20$, $N=2$, $T=3$ and $M_{IRS}=4$. In this case the order of complexity for the trellis based method is $4.1$ times that of the quantization method and $5\times 10^{-9}$ times that of the exhaustive search algorithm.  }

\section{IRS-Aided OMA} \label{Section IRS-Aided OMA}
\textcolor{black}{In this section we consider IRS-aided FDMA and TDMA communication systems  to provide a comparison benchmark for IRS-aided NOMA.
Here, as in the IRS-NOMA scenario, it is assumed that the acquired CSI is with an estimation error with the power of $\sigma_h^2$. 
In the following, the FDMA and TDMA cases are considered and a robust solution is given that maximizes their sum-rates. }
\textcolor{black}{\subsection{FDMA Mode}}
\textcolor{black}{In the FDMA mode, the AP communicates with the two users over two adjacent frequency resource blocks. In this case,  the achievable sum-rate is calculated by 
\begin{equation}\label{fdma opt prob}
    R^{[sum]}_{\text{FDMA}}=\frac{1}{2}\sum_{k=1}^2\log_2 \left(1+\frac{| {\hat{\mb{h}}}^{[k]}\mb{w}_{k}|^2}{\frac{1}{2}\sigma_h^2||\mb{w}_k||^2+\frac{1}{2}\sigma_{n}^2}\right).
\end{equation}
% \begin{equation}\label{fdma opt prob}
%     R^{[sum]}_{\text{OMA}}=\frac{1}{2}\sum_{k=1}^2\log_2 \left(1+\frac{| {\hat{\mb{g}}}^{[k]}\mb{v}_{k}|^2}{\frac{1}{2}\sigma_g^2||\mb{v}_k||^2+\frac{1}{2}\sigma_{w}^2}\right).
% \end{equation}
The proof can be easily driven with a similar approach to that  of theorem 1 in \cite{Omiddd1}. 
Therefore, the resultant optimization problem of maximizing the sum-rate is written as
\begin{equation}\label{main opt FDMA}
\begin{array}{cl}
\max\limits_{\mb{W}, \mb{v}}& R^{[sum]}_{\text{FDMA}}\\
\st 
&|\theta_i|=1, i=1,...,M,\\
&\tr \left(\mb{W}\mb{W}^H\right)\leq P.
\end{array}
\end{equation}
% \begin{equation}\label{main opt FDMA}
% \begin{array}{cl}
% \max\limits_{\mb{W}, \mb{v}}& R^{[sum]}_{\text{OMA}}\\
% \st 
% &|\psi_i|=1, i=1,...,M,\\
% &\tr \left(\mb{V}\mb{V}^H\right)\leq P_T.
% \end{array}
% \end{equation}
To solve this optimization problem we employ the alternating optimization (AO) technique, where in each step the problem is solved for each optimization variable separately until some convergence criteria is met.
Assuming that the transmit power on the $k$-th sub-channel is defined by $\mb{w}_k^H\mb{w}_k=P_k$, the optimization problem is presented by  
\begin{equation}\label{equi opt FDMA}
\begin{array}{cl}
\max\limits_{\mb{v},P_k,\mb{W}}& \sum_{k=1}^2\log_2 \left(1+\frac{| {\hat{\mb{h}}}^{[k]}\mb{w}_k|^2}{\frac{1}{2}P_k\sigma_h^2+\frac{1}{2}\sigma_{n}^2}\right)\\
\st 
&|\theta_i|=1, i=1,...,M,\\
&\sum_{k=1}^{K}P_{{k}}\leq P\\
&P_{{k}}\geq 0, \ k=1,\dots K,\\
&\mb{w}_{k}^{H}\mb{w}_{k}=P_{K},  \ k=1,\dots K.
\end{array}
\end{equation} }

\textcolor{black}{At first, we solve the optimization problem with respect to only $\mb{W}$. To do so, the optimization problem   can be solved separately for each $\mb{w}_k$. Since the logarithm is monotonically increasing with $|{\hat{\mb{h}}}^{[k]}\mb{w}_k|^2$, the solution can be driven by solving the following optimization problem 
\begin{equation} \label{FDMA wrt W}
\begin{array}{cl}
\max\limits_{\mb{w}_k}& |{\hat{\mb{h}}}^{[k]}\mb{w}_k|^2\\
\st 
&\mb{w}_{k}^{H}\mb{w}_{k}=P_{K},  \ k=1,\dots K.
\end{array}
\end{equation}
The closed-form solution to the optimization problem in (\ref{FDMA wrt W}) is given by 
\begin{equation}\label{FDMA solution to W}
\mb{w}_k = \sqrt{P_k}\frac{\hat{\mb{h}}^{[k]H}}{||\hat{\mb{h}}^{[k]}||}.
\end{equation}
We then rewrite the optimization problem in (\ref{equi opt FDMA}) by having (\ref{FDMA solution to W}) as
\begin{equation}\label{equi opt 2 FDMA}
\begin{array}{cl}
\max\limits_{\mb{v},P_k}& \sum_{k=1}^2\log_2 \left(1+\frac{P_k|| {\hat{\mb{h}}}^{[k]}||^2}{\frac{1}{2}P_k\sigma_h^2+\frac{1}{2}\sigma_{n}^2}\right)\\
\st 
&|\theta_i|=1, i=1,...,M,\\
&\sum_{k=1}^{K}P_{{k}}\leq P\\
&P_{{k}}\geq 0, \ k=1,\dots K.
\end{array}
\end{equation}}

\textcolor{black}{Next, we solve the optimization problem in (\ref{equi opt 2 FDMA}) with respect to the optimization variable $\mb{v}$. To do so, the  variable $P_k$ is considered as constant and since the objective function in (\ref{equi opt 2 FDMA}) is monotonically increasing with $\hat{\mb{h}}^{[k]}$, the problem is rewritten as 
\begin{equation}\label{Pt eq fdma}
\begin{array}{cl}
\max\limits_{P_{k},\mb{v}}& \sum_{k=1}^K P_{{k}}|| {\hat{\mb{h}}}^{[k]}||^2\\
\st 
&|\theta_i|=1, i=1,...,M.\\
\end{array}
\end{equation}
Then, for each phase shift value $\theta_i$, the optimization problem in (\ref{Pt eq fdma}) becomes 
\begin{equation}
    \begin{array}{cl}
    \max\limits_{|\theta_i|=1}& |\theta_i|^2{a}_{ii}^{[k]}+2\re\{(\sum_{j\neq i}^M\theta_j{a}_{ji}+{c}_i)\theta_i^*\},
    \end{array}
\end{equation}
where $a_{ij}$ and $c_i$, $i,j=1,...,M$, are the elements of $\mb{A}=\sum_{k=1}^K\hat{\mb{H}}_c^{[k]}\hat{\mb{H}}_c^{[k]H}$ and $\mb{c}=\sum_{k=1}^K\hat{\mb{h}}_{AU}^{[k]}\hat{\mb{H}}_c^{[k]H}$, respectively. Henceforth, the problem is solved alternately. In the $i$th step considering all other phase shift elements to be known, the optimal phase shift value for the $i$th element would be calculated as
\begin{equation}
    \theta_i=\frac{\sum_{j\neq i}^M \theta_j {a}_{ji}+{c}_i}{|\sum_{j\neq i}^M \theta_j {a}_{ji}+{c}_i|}.
\end{equation}
The optimization problem in (\ref{equi opt 2 FDMA}) with respect to the variable $P_k, k=1,2$ can be rewritten as
\begin{equation} \label{equi opt 3 FDMA}
\begin{array}{cl}
\max\limits_{P_1,P^{'}}& \log_2 \left(1+\frac{P_1|| {\hat{\mb{h}}}^{[1]}||^2}{\frac{1}{2}P_1\sigma_h^2+\frac{1}{2}\sigma_{n}^2}\right)+\log_2\left(1+\frac{(P^{'}-P_1)|| {\hat{\mb{h}}}^{[2]}||^2}{\frac{1}{2}(P^{'}-P_1)\sigma_h^2+\frac{1}{2}\sigma_{n}^2}\right) \\
\st 
&P^{'}\leq P,\\
&P_{{1}}\geq 0,
\end{array}    
\end{equation}
in which the variable $P_2$ is replaced by $P_2=P^{'}-P_1$. By calculating the derivative of the objective function in (\ref{equi opt 3 FDMA}) with respect to $P^{'}$, it can be seen that the derivative is always positive, and thus it is obvious that the objective function of the problem in (\ref{equi opt 3 FDMA}) is monotonically increasing with $P^{'}$. Now by taking the constraint $P^{'}\leq P$ into account, we can conclude that the optimal solution for this variable is $P^{'}=P$.
This can be readily proven by driving the derivative of the objective function of (\ref{equi opt 3 FDMA}) with respect to $P^{'}$. Thus, the problem is further simplified as 
\begin{equation} \label{equi opt 4 FDMA}
\begin{array}{cl}
\max\limits_{P_1}& \log_2 \left(1+\frac{P_1|| {\hat{\mb{h}}}^{[1]}||^2}{\frac{1}{2}P_1\sigma_h^2+\frac{1}{2}\sigma_{n}^2}\right)+\log_2\left(1+\frac{(P-P_1)|| {\hat{\mb{h}}}^{[2]}||^2}{\frac{1}{2}(P-P_1)\sigma_h^2+\frac{1}{2}\sigma_{n}^2}\right) \\
\st 
& 0 \leq P_{{1}}\leq P.
\end{array}    
\end{equation}
By driving the derivative of the objective function in (\ref{equi opt 4 FDMA}) with respect to $P_1$ and based on the first-order optimality condition, it can be seen that  only one optimal solution for $P_1$ exists, which can be found by  
\begin{equation}\label{P_1 solution}
    P_1 = \frac{-B+\sqrt{B^2-4AC}}{2A},
\end{equation}
where
\begin{equation}
    A = \frac{1}{4}\sigma_h^4(||\hat{\mb{h}}^{[1]}||^2+||\hat{\mb{h}}^{[2]}||^2),
\end{equation}
\begin{align}
    B =& \frac{1}{2}||\hat{\mb{h}}^{[2]}||^2\sigma_h^2\sigma_n^2+\sigma_n^2||\hat{\mb{h}}^{[1]}||^2||\hat{\mb{h}}^{[2]}||^2+\frac{P}{2}\sigma_h^4||\hat{\mb{h}}^{[1]}||^2\nonumber\\&+P\sigma_h^2||\hat{\mb{h}}^{[1]}||^2||\hat{\mb{h}}^{[2]}||^2+\frac{1}{2}||\hat{\mb{h}}^{[1]}||^2\sigma_h^2\sigma_n^2,
\end{align}
\begin{align}
    C =& \frac{1}{4}\sigma_n^4||\hat{\mb{h}}^{[2]}||^2-\frac{1}{4}\sigma_n^4||\hat{\mb{h}}^{[1]}||^2-\frac{P^2}{2}\sigma_h^2||\hat{\mb{h}}^{[1]}||^2||\hat{\mb{h}}^{[2]}||^2 \nonumber\\& -\frac{P}{2}\sigma_n^2\sigma_h^2||\hat{\mb{h}}^{[1]}||^2-\frac{P^2}{4}\sigma_h^4||\hat{\mb{h}}^{[1]}||^2-\frac{P}{2}\sigma_n^2||\hat{\mb{h}}^{[1]}||^2||\hat{\mb{h}}^{[2]}||^2
\end{align}
In case the solution given in (\ref{P_1 solution}) is within $[0,P]$, then $P_2$ is calculated by
\begin{equation}\label{P_2 solution}
    P_2=P-P_1.
\end{equation}
If not, $P_1$ is mapped to either $P$ or $0$ depending on whether $P_1>P$ or $P_1<0$, and based on the mapped value $P_2$ is calculated. 
It can be shown that the second-order derivative of the objective function in (\ref{equi opt 4 FDMA}) is always negative, thus the solution in (\ref{P_1 solution}) is the global maximum. }

\textcolor{black}{\subsection{TDMA Mode}}
\textcolor{black}{In the TDMA mode, the communication between the AP and the users happens in adjacent time-domain resource blocks. In this case, the achievable sum-rate is obtained by 
\begin{equation}
    R^{[sum]}_{\text{TDMA}}=\frac{1}{2}\sum_{k=1}^K\log_2 \left(1+\frac{| {\hat{\mb{h}}}^{[k]}\mb{w}_{k}|^2}{\frac{1}{2}\sigma_h^2||\mb{w}_k||^2+\frac{1}{2}\sigma_{n}^2}\right).
\end{equation}
Hence, the optimization problem becomes
\begin{equation}\label{TDMA main opt}
\begin{array}{cl}
\max\limits_{\mb{W}, \mb{v}}& R^{[sum]}_{\text{TDMA}}\\
\st 
&|\theta_i|=1, i=1,...,M,\\
&\tr \left(\mb{W}\mb{W}^H\right)\leq P.
\end{array}
\end{equation}
Similarly, we use the AO technique to decouple the variables and solve the problem in an iterative manner.  Considering $\mb{w}_k^H\mb{w}_k=P_k$, the optimization problem is rewritten the same way as (\ref{equi opt FDMA}). Similar to the FDMA mode, the problem with respect to the variable $\mb{w}_k$ is written by (\ref{FDMA wrt W}) which is solved by (\ref{FDMA solution to W}). Applying (\ref{FDMA solution to W}) to the problem in (\ref{FDMA wrt W}) results in the optimization problem in (\ref{equi opt 2 FDMA}) which needs to be solved once with respect to $\mb{v}$ and once with respect to $P_k, k=1,2$. }

% First we solve (\ref{equi opt 2 FDMA}) with respect to $\mb{v}$.
\textcolor{black}{Note that, unlike NOMA and FDMA, in TDMA the IRS phase shifts for each user can be optimized separately over different time slots. The IRS passive reflection can be time-selective, however it cannot be frequency-selective. With this fundamental difference in mind, the optimization problem in (\ref{equi opt 2 FDMA}) can be further simplified by decoupling to the following problems consecutively
\begin{equation}\label{psi eq tdma}
\begin{array}{cl}
\max\limits_{\mb{v}}& ||{\hat{\mb{h}}}^{[k]}||^2\\
\st 
&|\theta_i|=1, i=1,...,M.
\end{array}
\end{equation}
and
\begin{equation}\label{Pt eq tdma}
\begin{array}{cl}
\max\limits_{P_{k}}& \sum_{k=1}^K\log_2 \left(1+\frac{P_{{k}}|| {\hat{\mb{h}}}^{[k]}||^2}{\frac{1}{2}P_k\sigma_h^2+\frac{1}{2}\sigma_{n}^2}\right)\\
\st 
&\sum_{k=1}^{K}P_{{k}}\leq P\\
&P_{{k}}\geq 0, \ k=1,\dots K.
\end{array}
\end{equation}
To solve the problem in (\ref{psi eq tdma}) and obtaining the optimal phase shift vector for the $k$th sub-channel, with respect to only one specified $\theta_i, i=1,...,M$, the optimization problem can be written as
\begin{equation}
    \begin{array}{cl}
    \max\limits_{|\theta_i|=1}& |\theta_i|^2{a}_{ii}^{[k]}+2\re\{(\sum_{j=1}^{N}\hat{h}_{c_{ij}}^{[k]*}\hat{h}_{AU_j}^{[k]}+\sum_{j\neq i}^{M}\theta_j{a}_{ji}^{[k]})\theta_i^*\},
    \end{array}
    \end{equation}
which leads to an answer as
\begin{equation}
  \theta_i=\frac{\sum_{j=1}^{N}\hat{h}_{c_{ij}}^{[k]*}\hat{h}_{AU_j}^{[k]}+\sum_{j\neq i}^{M}\theta_j{a}_{ji}^{[k]}}{|\sum_{j=1}^{N}\hat{h}_{c_{ij}}^{[k]*}\hat{h}_{AU_j}^{[k]}+\sum_{j\neq i}^{M}\theta_j{a}_{ji}^{[k]}|},  
\end{equation}
where $a_{ij}$, $\hat{h}_{c_{ij}}^{[k]}$ and $\hat{h}_{AU_j}^{[k]}$, $i=1,...,M$, $j=1,...,N$, are the elements of $\mb{A}^{[k]}=\hat{\mb{H}}_c^{[k]}\hat{\mb{H}}_c^{[k] H}$, $\hat{\mb{H}}_{c}^{[k]}$ and $\hat{\mb{h}}_{AU}^{[k]}$, respectively.
As for the optimal value of $P_{K}$, it can be readily calculated via (\ref{P_1 solution}) and (\ref{P_2 solution}).}

\section{Numerical Results}\label{Section Numerical Results}
In this section, the performance of our proposed algorithm is evaluated in terms of the average sum-rate of the system.%, and the complexity of the proposed method is calculated and compared with other techniques in the literature. 
% \subsection{Performance Analysis}
The system model contains $K=2$ single-antenna users, an AP with $N=2$ active antennas and an IRS with $M$ antennas assists the communication between the AP and the users, where the value of $M$  is different in each simulation.  The large-scale fading coefficients, $\beta_{AU}$, $\beta_{IU}$ and $\beta_{AI}$ are modeled by 3GPP standards in \cite{3GPP2011}, where each large scale fading coefficient is calculated by $10\log_{10}(\beta)=-127.8-27\log_{10}(d)+Z$. In this model, $d$ represents the distance between the two nodes in km and the parameter $Z$ is a random variable with the distribution $Z\sim\mathcal{CN}(0,\sigma_{shad}^2)$ where $\sigma_{shad}^2=8 $ dB represents the shadowing. In general, it is assumed that the noise variance at each user is $\sigma_n^2=1$, the transmission power of the AP is $P=1$, and the estimation error variances are $\sigma^2=\sigma_{AU}^2=\sigma_{IU}^2=-10$ dB. 
% Without loss of generality, we assume that the IRS is placed near the edge cell, one user is randomly moving in a close range to the IRS while the other is randomly moving in the vicinity of the AP.
The simulation parameters for Algorithm. 1 are $\zeta=0.7$, $\eta=0.1$, $\epsilon=0.001$, $\rho=2.0661$, and the dual Lagrangian variables are initialized by setting all of their values to $0.1$. The initialization of all other variables in step zero is as follows, $\mb{W}^{0}=0.9\sqrt{(-1)\frac{P}{2KN}}\mb{1}$, $\mb{v}^{0}=\mb{1}$ and $\mb{a}^{0}=[0.5,0.5]^T$, where $\mb{1}$ is an all-one matrix/vector with the appropriate size.

\begin{figure}[t]
\begin{center}
  \includegraphics[scale=.30]{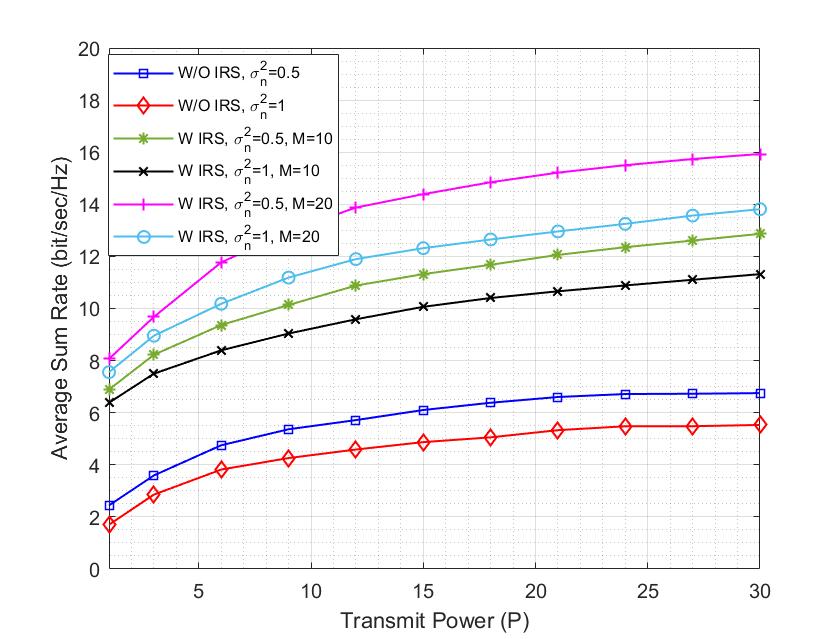}
      \caption{Average sum-rate vs. $P$; Comparison between NOMA with and without IRS in a perfect CSI scenario.}
    \label{fig:0}
\end{center}
\end{figure}

Fig. \ref{fig:0} depicts the impact of using IRS in a NOMA communication system. In this figure, the average sum rate is demonstrated versus the transmit power, for a NOMA system without IRS and two NOMA systems with continuous IRSs, one with $M=10$ and the other with $M=20$.  It is shown that even with small-scale IRS with only a few tens of antennas, the system performance is much better than when IRS is not employed. 
% Moreover, in terms of power efficiency, it can be claimed that IRS-NOMA is more efficient than NOMA without IRS, since in IRS-NOMA we would require less transmit power to achieve a certain sum-rate.
% \textcolor{\inquirycolor}{(It might be the case in general, but this statement doesn't stem from our system model and problem formulation. That is, rather than minimizing the transmission power, our optimization problem aims to optimize the sum-rate while maintaining the transmission power below a certain threshold. To judge about the power efficiency, we should manipulate the objective function according to a power transmission policy, which is out of the scope of this paper.) }
The advantages of IRS-NOMA are due to the capability of the IRS to modify the propagation environment such that the direction of users' channel vectors become more alike. 

\begin{figure}[t]
\begin{center}
  \includegraphics[scale=.30]{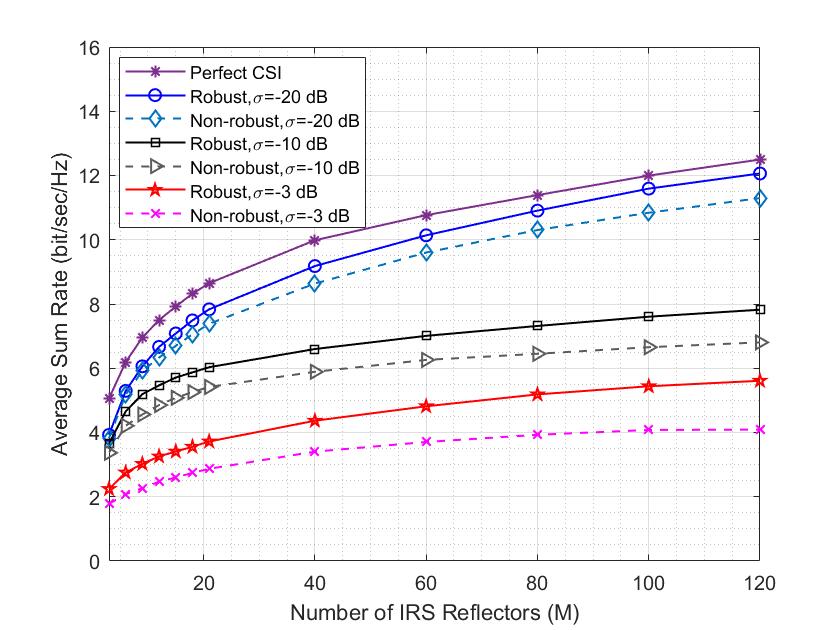}
      \caption{Average sum-rate vs. $M$; Comparison among robust, non-robust and perfect CSI designs.}
    \label{fig:1}
\end{center}
\end{figure}

Fig. \ref{fig:1} illustrates  the robustness of the proposed scheme to channel estimation error. The sum rate is depicted versus the number of IRS elements ($M$) for three cases of robust design, non-robust design and perfect CSI. In the non-robust design, it is assumed that the value of $\sigma^2=0$, hence we can deploy this scenario by considering the value of $\sigma_{h}^2$ in the design process to be zero, while channel estimation error exists{, i. e., $\sigma_{AU}^2$ and $\sigma_{BU}^2$ are non-zero,}  and it effects the output rate. 
It is shown that the performance of the proposed robust method is better than the non-robust case. Note that, as $M$ increases, the gap between the robust design and the non-robust structure widens, which is  a result of the approximation in (\ref{distribution of gtilda}), since the approximation in (\ref{distribution of gtilda}) is accurate when $M$ is large \cite{Omiddd1}. Moreover, as the channel estimation error increases, the performance gain of the robust design over non-robust system becomes more obvious.

\begin{figure}
    \centering
    \includegraphics[scale=0.30]{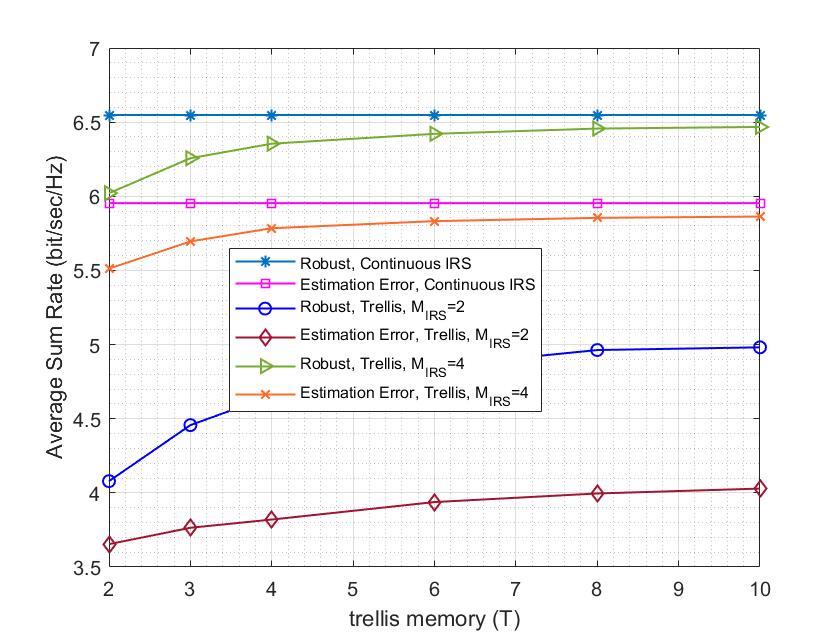}
    \caption{The average sum-rate vs. $T$, for $M=40$ and $M_{IRS}=[2,4]$.}
    \label{fig:rate_T}
\end{figure}

\textcolor{black}{In Fig \ref{fig:rate_T}, it is attempted to show the effectiveness of the trellis method for different IRS phase resolutions ($M_{IRS}$) and different trellis memories ($T$). In this figure, the average sum-rate is shown versus $T$  for an IRS with $M=40$ elements, with phase resolutions of $M_{IRS}=[2,4]$. It is shown that the performance of the trellis method in case $M_{IRS}=4$ is very close to the performance of continuous IRS-NOMA, specially at $T\geq 3$. Thus, for the rest of the simulations, we set the values $M_{IRS}=4$ and $T=3$, to guarantee a near-optimal performance and ensure that the computational complexity of the method is low.}

\begin{figure}[t]
\begin{center}
  \includegraphics[scale=.58]{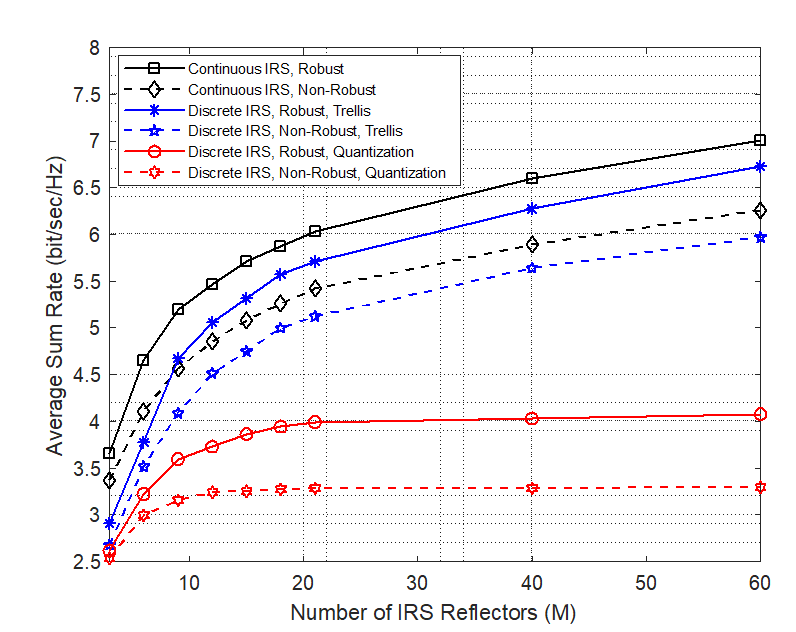}
      \caption{Average sum-rate vs. $M$; Comparison among robust, non-robust designs in $\sigma=-10 dB$, for two cases of continuous and discrete phase shifters in the IRS.}
    \label{fig:2}
\end{center}
\end{figure}

\textcolor{black}{In Fig. \ref{fig:2}, the performance of the trellis-based discrete phase shift selection method is compared to the continuous case. It is assumed that the channel estimation error variance is $\sigma^2=-10$  dB, and the performance of both robust and non-robust cases are given with continuous and discrete IRS phase shifts. The trellis parameters are $T=3$ and $M_{IRS}=4$. 
% Needless to say, by increasing the memory $T$ and the resolution of the phase shifter $M_{IRS}$, the performance would be better, but the complexity would be higher as well, specially for large values of $M$.
As depicted here, the performance of a 2-bit discrete IRS using the trellis method is very much close to that of the continuous IRS. Also it is shown that using the trellis method for a discrete IRS out-performs the quantization technique. Comparing the performance of the trellis-based method with the quantization method in this figure, we can see that the out-performance of the trellis method comes at the cost of a slight rise in complexity, which can be considered negligible (refer to section  \ref{Discrete IRS}). }

\begin{figure}[t]
\begin{center}
  \includegraphics[scale=.30]{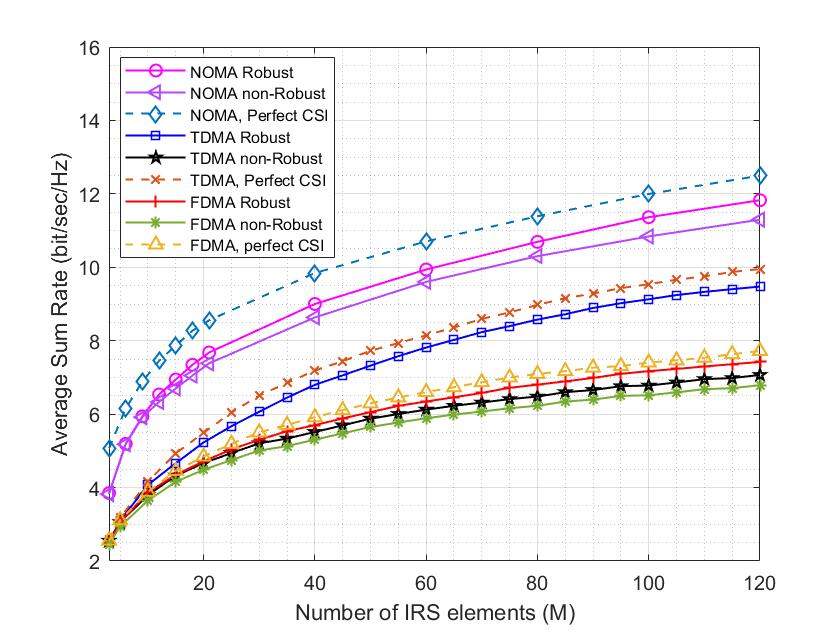}
      \caption{Average sum-rate vs. $M$; Comparison among IRS-aided FDMA, TDMA and NOMA.}
    \label{fig:3}
\end{center}
\end{figure}

\textcolor{black}{Fig. \ref{fig:3}, proposes a comparison between IRS-assisted NOMA and OMA systems where $K=2$ users are served by an AP with $N=2$ antennas. In this figure, the average sum rate is shown versus the number of IRS elements, for $\sigma=-20$dB, $P=1$ and $\sigma_n=1$. 
In this figure, it is demonstrated how IRS-aided NOMA outperforms OMA in terms of average sum-rate. It is noteworthy that, in case non-overlapping resource blocks  are dedicated to users, the throughput of NOMA should be theoretically twice that of OMA thanks to employing twice as much resource block as OMA, and IRS is a beneficial tool for realizing such a situation. However, in practice, there would be some residual co-channel interference due to non-ideal similarity in the directions of users’ channels. 
% As the number of reflectors rises, the interference could be further suppressed by applying IRS to NOMA, and this direction similarity tends to the ideal case; hence, the achievable rate would be almost twice that of OMA in larger number of reflectors. 
It is also shown that TDMA has a superior performance than FDMA. This is driven by the fact that in TDMA, the IRS phase shift matrix is adjusted for each user separately, while in FDMA the phase shift matrix  should be in charge of optimizing both users' propagation environments at the same time, making the IRS less flexible to the channel variation. In other words, FDMA is not able to fully leverage the degree of freedom rendered by  IRS.}
% \textcolor{red}{Note that, as we relax the relevant optimization problem to some extent in FDMA case, the system suffers form negligible performance loss.}

\begin{figure}[t]
\begin{center}
  \includegraphics[scale=.30]{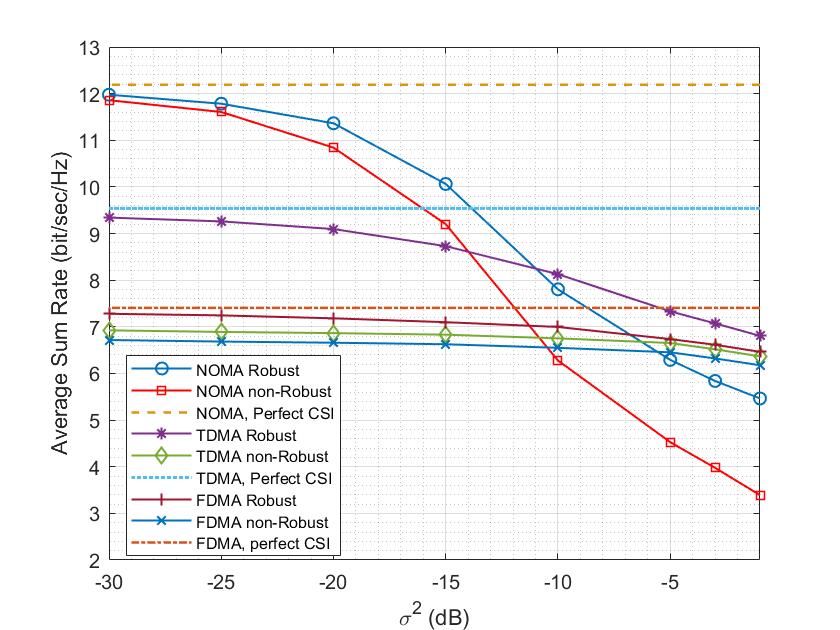}
      \caption{Average sum-rate vs. $\sigma^2$; Comparison among IRS-aided FDMA, TDMA and NOMA.}
    \label{fig:OMA_NOMA_sigma}
\end{center}
\end{figure}

\textcolor{black}{In Fig. \ref{fig:OMA_NOMA_sigma} the average sum rate is shown versus different values of channel estimation error variance $\sigma^2$. This figure compares the performance of IRS-NOMA with IRS-OMA in three cases of perfect CSI, imperfect CSI with robust design and imperfect CSI with non-robust design. As shown in this figure, in the perfect CSI scenario IRS-NOMA  always renders higher performance compared to IRS-OMA due to channel reconfiguration abilities of the IRS. However, This promising feature of the IRS can hardly extend to the case undergoing severe channel uncertainty. This limitation stems from the fact that for a high channel uncertainty level, the system operates in a noise-dominant environment, making both the NOMA and OMA methods inefficient.  It is shown that while the robust IRS-NOMA design can compensate for the channel uncertainty up to a certain point,   at $\sigma^2>-11$ dB the performance of robust IRS-NOMA becomes worse than that of robust IRS-TDMA as a consequence of a noise-dominant environment. }
% \textcolor{red}{and at $\sigma^2>-7$dB it becomes even worse than robust IRS-FDMA. Hence, it can be concluded that the choice between IRS-NOMA or IRS-OMA depends highly upon the channel uncertainty.}

% In Fig. \ref{fig:4}, the performance of  IRS-aided NOMA is compared to FDMA and TDMA and it is evaluated how each system works in case of different values for the maximum transmit power ($P$). The system model contains $K=2$ users, the AP has $N=2$ antennas and the IRS has $M=20$ passive elements.  As depicted here, the IRS-aided NOMA outperforms OMA in terms of average sum rate. Also, it is noteworthy that by increasing the power limit, the gap between NOMA and OMA widens. This shows the power efficiency of IRS-aided NOMA compared to OMA. {In general, apart from the noise term, IRS-assisted NOMA encounters two interfering components, i.e., the co-channel interference between users and the attenuation of users’ channels, whereas IRS-aided OMA only faces the former due to utilizing separate sub-channels. Therefore, increasing the total power results in further rate improvement in the NOMA case, as the OMA performance saturates in lower power levels.}

% \begin{figure}[t]
% \begin{center}
%   \includegraphics[scale=.6]{fig4.eps}
%       \caption{Average sum-rate versus $P$; A comparison among IRS-aided FDMA, TDMA and NOMA systems in perfect CSI scenario.}
%     \label{fig:4}
% \end{center}
% \end{figure}

% \begin{table}[h] 
%     \centering
%     \begin{tabular}{c|c}
%         Step & Order of Complexity  \\ \hline
%         1 & $\mathcal{O}(32N+44)$  \\ \hline
%         2 & $\mathcal{O}(2N^3+16N^2+40N+40)$
%          \\ \hline
%         3 & $\mathcal{O}(53)$
%          \\ \hline
%         4 & $\mathcal{O}(16N+67 )$
%          \\ \hline
%         5 & $\mathcal{O}(110)$
%          \\ \hline
%         6 & $\mathcal{O}(4N^2+20N+6)$
%          \\ \hline
%         7 & $\mathcal{O}(4M^2+(2N^2+4N+5)M+N^2+7N+6)$
%          \\ \hline
%         8 & $\mathcal{O}(2M^2+(4N+4)M+4N)$
%          \\ \hline
%         9 & $\mathcal{O}(18N+60)$ \\ \hline
%         10 & $\mathcal{O}(20N+54)$
%     \end{tabular}
%     \caption{Order of complexity for each step in Algorithm. \ref{Alg22}}
%     \label{complexity table 1}
% \end{table}

\section{Conclusion}\label{Section Conclusion}
This paper investigates the benefit of utilizing IRS in a NOMA communication system. We attempted to jointly optimize the beamforming vector and the IRS entries under the assumption of imperfect CSI, to maximize the sum rate. The PDD algorithm is used to design a robust joint beamforming at the IRS and the AP. The proposed method is an iterative algorithm with closed-form solutions in each step, hence, the computational complexity is very low. \textcolor{black}{Two cases of continuous and discrete IRS are considered in this paper, and a trellis-based solution is proposed for the discrete phase shift selection, which is shown to out-perform the conventional methods.} Numerical results show the efficiency of the robust design compared to the non-robust and perfect CSI scenarios. To investigate the benefits of NOMA compared to OMA, two cases of IRS-assisted FDMA and TDMA systems are considered. It is shown  that IRS-aided NOMA outperforms OMA in terms of spectral efficiency when the channel estimation error is low. 

\appendices

\section{Proof of Theorem 1}\label{appendixB}
 Let $\mathcal{I}\left(\alpha_{i}\mb{x}^{[i]} ; y^{[i]} | \hat{\mb{h}}\right)$ be the conditional mutual information of user $i$ conditioned on estimated channel matrix $\hat{\mb{h}}$. Expanding $\mathcal{I}\left(\alpha_{i}\mb{x}^{[i]} ; y^{[i]} | \hat{\mb{h}}\right)$ in terms of the differential entropies results in
\begin{equation}\label{Mutual Information}
   \mathcal{I}\left(\alpha_{i}\mb{x}^{[i]} ; y^{[i]} | \hat{\mb{h}}\right)=\mathcal{H}\left(\alpha_{i}\mb{x}^{[i]} | \hat{\mb{h}}\right)-\mathcal{H}\left(\alpha_{i}\mb{x}^{[i]} | y^{[i]}, \hat{\mb{h}}\right) 
\end{equation}
%Since $\mb{x}^{[i]}$ is jointly Gaussian, 
The first term on the right hand side of (\ref{Mutual Information}) simplifies
to $ \log_2 \det\left(2 \pi e \mb{v}_{i}\right),$ where $\mb{v}_{i} \triangleq \mathbb{E}\left\{\alpha_{i}^2\mb{x}^{[i]} \mb{x}^{[i] {H}}\right\}$ denotes the
transmit co-variance matrix related to $\mb{x}^{[i]}$ \cite{Cover1991}. Regarding the equation in (\ref{received signal2}), the second term of the right hand side of (\ref{Mutual Information}) is upper bounded by the entropy of a Gaussian random variable \cite{Medard2000} as follows
\begin{align}\label{Entrop bound}
    &\mathcal{H}(\alpha_{i}\mb{x}^{[i]} | y^{[i]}, \hat{\mb{h}}) \leq  \log_2  \det\Big(2 \pi e(\mb{v}_{i}-\frac{\mb{v}_{i} \hat{\mb{h}}^{[i] H} \hat{\mb{h}}^{[i]} \mb{v}_{i}}{\hat{\mb{h}}^{[i]} \mb{v}_{i} \hat{\mb{h}}^{[i] H}+\Gamma_{i}})\Big),
\end{align}
in which
\begin{equation}
    \Gamma_{i}\triangleq \begin{cases}
     \sum_{k=1}^{K} \sigma_{h}^{2} \tr\left(\mb{v}_{k}\right)+\sigma_{n}^{2},& i=1
     \\ \\  \hat{\mb{h}}^{[2] H} \mb{v}_{1} \hat{\mb{h}}^{[2]}+\sum_{k=1}^{K} \sigma_{h}^{2} \tr\left(\mb{v}_{k}\right)+\sigma_{n}^{2}, & i=2.
     \end{cases}
    \end{equation}
Now, we employ the Woodbury matrix identity as follows
\begin{equation}\label{woodbury}
    (\mb{A}+\mb{B} \mb{C D})^{-1}=\mb{A}^{-1}-\mb{A}^{-1} \mb{B}\left(\mb{C}^{-1}+\mb{D} \mb{A}^{-1} \mb{B}\right)^{-1} \mb{D} \mb{A}^{-1}.
\end{equation}
Assuming $\mb{A}=\mb{I}$, $\mb{B}=\hat{\mb{h}}^{[i] H}$, $\mb{C}=\Gamma_{i}^{-1}$ and
$\mb{D}=\hat{\mb{h}}^{[i]} \mb{v}_{i},$ and using (\ref{woodbury}), it is concluded
\begin{equation} \label{Matrix identity}
    \mb{I}-\frac{\hat{\mb{h}}^{[i] H} \hat{\mb{h}}^{[i]} \mb{v}_{i}}{\hat{\mb{h}}^{[i]} \mb{v}_{i} \hat{\mb{h}}^{[i] H}+\Gamma_{i}}=\left(\mb{I}+\frac{\hat{\mb{h}}^{[i] H} \hat{\mb{h}}^{[i]} \mb{v}_{i}}{\Gamma_{i}}\right)^{-1}.
\end{equation}
\\
Thus, the right hand side of (\ref{Entrop bound}) can be rewritten
as
\begin{align}\label{Matrix identity}
 &\log_2 \det\left(2 \pi e \mb{v}_{i}\left(\mb{I}-\frac{\hat{\mb{h}}^{[i] H} \hat{\mb{h}}^{[i]} \mb{v}_{i}}{\hat{\mb{h}}^{[i]} \mb{v}_{i} \hat{\mb{h}}^{[i] H}+\Gamma_{i}}\right)\right)\nonumber\\&= \log_2 \det\left(2 \pi e \mb{v}_{i}\left(\mb{I}+\frac{\hat{\mb{h}}^{[i] H} \hat{\mb{h}}^{[i]} \mb{v}_{i}}{\Gamma_{i}}\right)^{-1}\right)\nonumber \\
&= \log_2 \det\left(2 \pi e \mb{v}_{i}\right)- \log_2 \det\left(\mb{I}+\frac{\hat{\mb{h}}^{[i] H} \hat{\mb{h}}^{[i]} \mb{v}_{i}}{\Gamma_{i}}\right).
\end{align}
\\
Exploiting (\ref{Matrix identity}) and employing Sylvester's determinant theorem, i. e.,  $\det(\mb{I}+\mb{A B})=\det(\mb{I}+$
BA),  (\ref{Entrop bound}) is rewritten by
\begin{align}\label{simplified entropy}
\mathcal{H}\left(\alpha_{i}\mb{x}^{[i]} | y^{[i]}, \hat{\mb{h}}\right) & \leq \log_2 \det\left(2 \pi e \mb{v}_{i}\left(\mb{I}-\frac{\hat{\mb{h}}^{[i] H} \hat{\mb{h}}^{[i]} \mb{v}_{i}}{\hat{\mb{h}}^{[i]} \mb{v}_{i} \hat{\mb{h}}^{[i] H}+\Gamma_{i}}\right)\right)\nonumber \\
&= \log_2 \det\left(2 \pi e \mb{v}_{i}\right) - \log_2 \det\left(1+\frac{\hat{\mb{h}}^{[i]} \mb{v}_{i} \hat{\mb{h}}^{[i] H}}{\Gamma_{i}}\right).
\end{align}
Consequently, assuming $\mathbb{E}\left\{\left|s^{[i]}\right|^{2}\right\}=1$ and utilizing (\ref{simplified entropy}),  (\ref{Mutual Information}) leads to
\begin{align}
    \mathcal{I}\left(\alpha_{i}\mb{x}^{[i]} ; y^{[i]} | \hat{\mb{h}}\right) &\geq  \log_2 \det\left(1+\frac{\hat{\mb{h}}^{[i]} \mb{v}_{i} \hat{\mb{h}}^{[i] H}}{\Gamma_{i}}\right)= \log_2 \left(1+\frac{\left|\alpha_{i}\hat{\mb{h}}^{[i]}\mb{w}_{i}\right|^{2}}{\Gamma_{i}}\right).
\end{align}
\\
Therefore, the minimum achievable rate of the $i$-th user is written by 
\begin{equation}\label{user rate}
    R^{[i]}=\begin{cases}
    \log_2 \left(1+\frac{\alpha_{1}^{2}|\hat{\mb{h}}^{[1]}\mb{w}_{1}|^2}{\sigma_{h}^{2}\left(||\mb{w}_{1}||^2+||\mb{w}_{2}||^2\right)+\sigma_{n}^2}\right),& i=1\\ & \\
    \log_2 \left(1+\frac{\alpha_{2}^{2}|\hat{\mb{h}}^{[2]}\mb{w}_{2}|^2}{\alpha_{1}^{2}|\hat{\mb{h}}^{[2]}\mb{w}_{1}|^2+\sigma_{h}^{2}\left(||\mb{w}_{1}||^2+||\mb{w}_{2}||^2\right)+\sigma_{n}^2}\right),& i=2.
    \end{cases}
    \end{equation}
    
    % \section{Trellis Algorithm}\label{trellis appendix}
\bibliographystyle{IEEEtran}
\bibliography{Ref1}
\end{document}